\documentclass[12pt]{article}

\font\grande=cmr9.5 scaled \magstep4
\font\medio=cmr9.5 scaled \magstep2
\outer\def\beginsection#1\par{\medbreak\bigskip
      \message{#1}\leftline{\bf#1}\nobreak\medskip
\vskip-\parskip
      \noindent}
\usepackage{graphicx} 
\begin{document}
\bibliographystyle {unsrt}

\titlepage

\begin{flushright}
CERN-PH-TH/2015-182
\end{flushright}

\vspace{10mm}
\begin{center}
{\grande The spectrum of anomalous magnetohydrodynamics}\\
\vspace{1.5cm}
 Massimo Giovannini
 \footnote{Electronic address: massimo.giovannini@cern.ch}\\
\vspace{1cm}
{{\sl Department of Physics, 
Theory Division, CERN, 1211 Geneva 23, Switzerland }}\\
\vspace{0.5cm}
{{\sl INFN, Section of Milan-Bicocca, 20126 Milan, Italy}}
\vspace*{0.5cm}
\end{center}

\vskip 0.5cm
\centerline{\medio  Abstract}
The equations of anomalous magnetohydrodynamics describe an Abelian plasma 
where conduction and chiral currents are simultaneously present and constrained by the second law of thermodynamics. At high frequencies  the magnetic currents play the leading role and the spectrum is dominated by  two-fluid effects. The system behaves instead as a single fluid in the low-frequency regime where the vortical currents  induce potentially large hypermagnetic fields. After deriving the physical solutions of the generalized Appleton-Hartree equation,  the corresponding dispersion relations 
are scrutinized and compared with the results valid for cold plasmas.  Hypermagnetic knots and fluid vortices can be concurrently present at very low frequencies and suggest 
a qualitatively different dynamics of the hydromagnetic nonlinearities. 
\vskip 0.5cm

\noindent

\vspace{5mm}

\vfill
\newpage
\renewcommand{\theequation}{1.\arabic{equation}}
\setcounter{equation}{0}
\section{Introduction}
\label{sec1}
Electrically conducting media are customarily described as a single fluid  in the low-frequency branch of the plasma spectrum. This approach has been extensively applied to the analysis of hydromagnetic nonlinearities  \cite{moffat} evolving in terrestrial \cite{biskamp} and astrophysical plasmas \cite{alf,parker,zeldovich}. The same strategy cannot be extended to higher frequencies where the one-fluid description is no longer tenable \cite{stix} and the plasma must be treated, at least, as a double fluid. This well known aspect of conventional electromagnetic plasmas stems directly from the properties of the vector currents which are associated, in the high-frequency limit,  with the ions and with the electrons. When the plasma is globally neutral  the total vector current is instead Ohmic in the low-frequency domain.

A problem of similar nature occurs in anomalous magnetohydrodynamics \cite{mg2013} describing a charged fluid where axial and vector currents are simultaneously present: while the axial currents are not conserved because of the triangle anomaly, the vector currents are eventually Ohmic.  The purpose of this investigation is a systematic discussion of the spectrum of anomalous magnetohydrodynamics (AMHD in what follows). The equations of AMHD differ from the ones where only chiral currents are present \cite{rub,red} at finite fermionic density. They generalize the system firstly explored in Ref. \cite{mg19982000} accounting for the evolution of the hypermagnetic and hyperelectric fields in the electroweak plasma. Indeed, in the symmetric phase of the electroweak theory the 
non-screened vector modes of the plasma correspond to the hypercharge which has a chiral coupling to fermions. 
 The axial currents may be associated with the evolution 
of the chemical potential or with the presence of an axionlike field \cite{axion1,axion2} (see also \cite{ax3}). 
In both cases the plasma may host parity-odd configurations of the gauge fields characterized
non-vanishing hypermagnetic gyrotropy $\vec{B} \cdot \vec{\nabla} \times \vec{B}$ which is 
the hydromagnetic analog of the kinetic gyrotropy (i.e. $\vec{v} \cdot \vec{\nabla} \times \vec{v}$) 
naturally appearing in the discussion of mean-field dynamos \cite{moffat,parker}. 
The dynamical production of hypermagnetic knots and Chern-Simons waves during inflation offers a potentially viable mechanism for the generation of the baryon asymmetry of the Universe (see last two papers in \cite{mg19982000}). In AMHD the hypermagnetic current is complemented by a vortical current possibly leading to the formation of fluid vortices.

The same class of physical systems previously discussed in the electroweak plasma 
also arises in the framework of the so called chiral magnetic effect \cite{kh}. Both phenomena are often presented as 
macroscopic manifestations of triangle anomalies. The model of chiral liquid emerging in the context of AMHD could then be 
relevant also in the context of the chiral magnetic effect insofar as axial currents and quark vector currents 
are concurrently present in the strongly interacting plasma. In the absence of finite conductivity effects (see e.g. \cite{hol,SS}) the validity of the second law of thermodynamics is guaranteed by the simultaneous presence of an hypermagnetic current and of a chiral vortical term. In AMHD the vector currents (eventually responsible of Ohmic dissipation), the chiral 
currents (determining the anomalous effects) and the vortical currents (required by the second principle 
of thermodynamics) are all described by the appropriate kinetic coefficients. 
Whenever possible AMHD will be discussed in analogy with the spectrum of conventional plasmas. More specifically the plan of this investigation is the following. In section \ref{sec2} we discuss the relativistic problem and derive the general form of the kinetic coefficients. In section \ref{sec3} the properties of the two-fluid equations are analyzed while section \ref{sec4} is devoted to the dispersion relations in the high-frequency domain.  The one-fluid equations and their implications are presented in section \ref{sec5}. Section \ref{sec6} contains our concluding remarks. To avoid digressions some relevant technical aspects have been relegated to the appendix.

\renewcommand{\theequation}{2.\arabic{equation}}
\setcounter{equation}{0}
\section{The relativistic formulation and the total entropy}
\label{sec2}
The conservation of the total energy momentum tensor and the evolution 
of the chiral and vector currents determine the relativistic form of the second law of thermodynamics. 
If the four-divergence of the entropy four-vector is to be positive semi-definite (as implied by the generalized second law) the chiral and vector currents must contain supplementary kinetic coefficients corresponding to the hypermagnetic and to the vortical currents. In what follows the dissipative effects are included in the framework of the Landau approach: the total four-velocity coincides then with the velocity of the energy transport defined from the mixed components of the total energy-momentum tensor.  

\subsection{Ohmic and chiral currents}
In the simplest situation the total energy-momentum tensor of the system ($T_{\mu\nu}^{(tot)}$ in what follows) 
consists of four qualitatively different contributions: 
the energy-momentum tensor of the charged species (denoted by $T_{\mu\nu}^{(\pm)}$), the energy-momentum tensor 
of the chiral species (labeled by $T_{\mu\nu}^{(R)}$), the dissipative contribution $T_{\mu\nu}^{(diss)}$ 
and the gauge contribution ${\mathcal T}_{\mu\nu}$ (corresponding to an hypercharge gauge-field strength $Y_{\mu\nu}$): 
\begin{equation}
T_{\mu\nu}^{(tot)} = T_{\mu\nu}^{(+)} + T_{\mu\nu}^{(-)} + T_{\mu\nu}^{(R)}+ T_{\mu\nu}^{(diss)} + {\mathcal T}_{\mu\nu}.
\label{enm1}
\end{equation}
The covariant conservation of $T_{\mu\nu}^{(tot)}$ implies, as usual, that $\nabla_{\mu} T^{\mu\nu}_{(tot)}=0$ where $\nabla_{\mu}$ denotes the covariant derivative\footnote{The discussion will be conducted in a general relativistic formulation 
even if the spectrum of AMHD will be discussed in flat space-time.} defined from the metric tensor $g_{\mu\nu}$ (with signature mostly minus). 
The chiral and the conduction currents coexist but are not bound to coincide: they obey different equations. More specifically 
 the anomalous current is not covariantly conserved and its evolution can be written as\footnote{Note that $\widetilde{n}_{R}$ and $u^{\mu}_{R}$ are respectively the concentration and the four-velocity of the chiral species.}
\begin{equation}
\nabla_{\mu} j^{\mu}_{R} = {\mathcal A}_{R} \, Y_{\alpha\beta} \widetilde{Y}^{\alpha\beta}, \qquad j^{\mu}_{R} = \widetilde{n}_{R} \, u^{\mu}_{R} + \nu^{\mu}_{R},
\label{enm2}
\end{equation}
where ${\mathcal A}_{R}$ is a numerical factor that is determined by the specific nature of the chiral species and by the coupling to the hypercharge field; note that in the Landau frame $\nu^{\mu}_{R} u_{\mu}^{R} =0$. Conversely the conduction current is covariantly conserved, it is a source of the evolution equations of the gauge fields and it may even contain a dissipative contribution:
\begin{equation}
\nabla_{\mu} \,Y^{\mu\nu} = 4 \pi j^{\nu}, \qquad j^{\nu} = j^{\nu}_{+} + j^{\nu}_{-}, \qquad \nabla_{\mu} \, j^{\mu} =0,
\label{enm3}
\end{equation}
where $j^{\nu}_{\pm} =(q_{\pm} \widetilde{n}_{\pm} \,\, u^{\mu}_{\pm} \,+ \, \nu_{\pm}^{\mu})$. The dual field strength $\widetilde{Y}^{\mu\nu}$ obeys, as usual, $\nabla_{\mu} \widetilde{Y}^{\mu\nu}=0$. This is the approach followed in \cite{mg2013} which 
differs from other more conventional approaches \cite{hol,SS} where the anomalous current and the conduction current are 
identified. In the present approach the anomalous current is not directly the source of the evolution of the hypercharge.

In a general relativistic description $Y_{\mu\nu}$ and its dual account for the evolution of the hypercharge field
\footnote{As soon as we speak of hyperelectric and hypermagnetic fields we are implicitly assuming that the plasma has a finite conduction current so that a preferred frame can be selected where the electric fields are suppressed. Even if the electric and magnetic fields are non-relativistic concepts,  it is practical to introduce the electric and the magnetic components of the gauge field strength in a generally covariant language.} 
however the gauge field strength can be decomposed into the hyperelectric and hypermagnetic parts denoted, respectively, by ${\mathcal E}^{\mu}$ and ${\mathcal B}^{\mu}$:
\begin{equation}
Y_{\alpha\beta} = {\mathcal E}_{\alpha}\, u_{\beta}\, - {\mathcal E}_{\beta} u_{\alpha}\, + \, E_{\alpha\beta\rho\sigma}\, u^{\rho} {\mathcal B}^{\sigma},
\label{C1}
\end{equation}
where $E_{\alpha\beta\rho\sigma} = \sqrt{-g}\, \epsilon_{\alpha\beta\rho\sigma}$ and
 $\epsilon_{\alpha\beta\rho\sigma}$ is the four-dimensional Levi-Civita symbol while $g$ is the determinant of the metric tensor.  The total four-velocity of the system follows from
\begin{equation}
(p+ \rho ) u^{\mu} \, u^{\nu} = \sum_{a} [p_{(a)} + \rho_{(a)}]\, u^{\mu}_{(a)} \, u^{\nu}_{(a)},
\label{th4}
\end{equation}
where $w = (\rho + p)$ denotes the total enthalpy; the sum in Eq. (\ref{th4}) runs over all the species of the plasma, both charged and chiral. Close to an  equilibrium situation the four-velocity of the anomalous species coincides with the bulk velocity of the plasma and, therefore, $u_{R}^{\mu} \simeq u^{\mu}$.  The vorticity four vector can then be defined as:
\begin{equation}
\omega^{\mu} = \widetilde{f}^{\mu\alpha} u_{\alpha} \equiv \frac{1}{2} E^{\mu\alpha\beta\gamma} \, u_{\alpha}\, f_{\beta\gamma}, \qquad f_{\beta\gamma} = \nabla_{\beta} u_{\gamma} - \nabla_{\gamma} u_{\beta}.
\label{C2}
\end{equation}
From Eqs. (\ref{C1}) and (\ref{C2}) it follows that the four-divergences of ${\mathcal E}^{\mu}$, ${\mathcal B}^{\mu}$ and $\omega^{\mu}$
are given by:
\begin{eqnarray}
w\,\nabla_{\mu} \omega^{\mu} &=& - 2 \omega^{\alpha} \partial_{\alpha} p  - 2 \widetilde{n} {\mathcal E}^{\alpha} \omega_{\alpha},
\label{C3}\\
w\, \nabla_{\mu} {\mathcal B}^{\mu} &=& 2 w\, Y_{\rho\sigma} \, \omega^{\rho}\, u^{\sigma} + u_{\mu}\, \partial_{\alpha} p \widetilde{Y}^{\mu\alpha} + u_{\mu} \, Y_{\alpha\beta} \, j^{\beta} \widetilde{Y}^{\mu\alpha},
\label{C4}\\
w\,\nabla_{\mu} {\mathcal E}^{\mu} &=& w [4 \pi j^{\alpha} u_{\alpha}  - \widetilde{Y}^{\mu\rho}\omega_{\mu} u_{\rho} ] + 
Y^{\beta\gamma} u_{\beta} \partial_{\gamma} p + Y^{\beta\gamma}u_{\beta} Y_{\gamma\alpha} j^{\alpha},
\label{C5}
\end{eqnarray}
where $w = p + \rho$ is the enthalpy density of the fluid. Equations (\ref{C3}), (\ref{C4}) and (\ref{C5}) have been obtained 
in the globally neutral case where $\widetilde{n} = \widetilde{n}_{+} = \widetilde{n}_{-}$ and  $q_{+} = q = - q_{-}$ but they 
can be easily generalized to the case where the plasma is not globally neutral.

\subsection{First and second principles of thermodynamics}
Denoting with $\mu_{R}$ the chemical potential associated with the anomalous species, the first principle of thermodynamics demands: 
\begin{equation}
d E = T d S - p d V + \mu_{R} d N_{R},\qquad w=\rho + p = T \varsigma + \mu_{R}\, \widetilde{n}_{R}.
\label{th1}
\end{equation}
The fundamental identity  $E = T S - p V + \mu_{R} N_{R}$ can be divided by a fiducial volume and the result is the one reported in the second relation of Eq. (\ref{th1}) 
where $\varsigma$ is the entropy density and $\rho$ the total energy density of the system. Combining the two relations 
of Eq. (\ref{th1}) further thermodynamic relations can be obtained\footnote{Like, for instance, $\varsigma \partial_{\alpha} T + \widetilde{n}_{R} \partial_{\alpha} \mu_{R} = \partial_{\alpha} p$ or $\partial_{\alpha} \rho = T \partial_{\alpha} \varsigma + \mu_{R} \partial_{\alpha} \widetilde{n}_{R}$.}.
Since the anomaly-induced currents are protected by topology they are not associated with dissipative effects. 
Thus the entropy production of the plasma must only come, in the relativistic case, from the viscosity coefficients 
or from the Ohmic contributions but neither from the chiral currents nor from the corresponding diffusive contribution. 
The absence of dissipative contributions stemming from the anomalous sector
demands that the total entropy four-vector must be supplemented by two further coefficients ${\mathcal S}_{\omega}$ and 
${\mathcal S}_{B}$:
\begin{equation}
\varsigma^{\mu} = \varsigma u^{\mu} - \overline{\mu}_{R} \nu_{R}^{\mu} + {\mathcal S}_{\omega} \omega^{\mu} + {\mathcal S}_{B} {\mathcal B}^{\mu}.
\label{th2}
\end{equation}
The covariant conservation of the total energy momentum tensor $T^{(tot)}_{\mu\nu}$ can be 
written as 
\begin{equation}
\nabla_{\mu} \varsigma^{\mu} - \frac{\sigma}{T} Y^{\alpha\beta} Y_{\nu\alpha} u^{\nu} u_{\beta} - \frac{T^{\mu\nu}_{(diss)}}{T} \nabla_{\mu} u_{\nu} = {\mathcal Z},
\label{th5}
\end{equation}
where we assumed, for the sake of simplicity, a global charge neutrality of the plasma and a
corresponding Ohmic form for the charged species, namely
${\mathcal P}^{\alpha}_{\mu} \, j^{\mu} = \sigma Y^{\alpha\nu} \, u_{\nu}$ where  ${\mathcal P}^{\alpha}_{\mu}= \delta_{\mu}^{\alpha} - u_{\mu} u^{\alpha}$ is the 
standard projector.
The function ${\mathcal Z}$ appearing in Eq. (\ref{th5}) is given by
\begin{equation}
{\mathcal Z} =  \nabla_{\mu}\biggl( {\mathcal S}_{\omega} \, \omega^{\mu}  + {\mathcal S}_{B} \,{\mathcal B}^{\mu}\biggr)
- \frac{\nu^{\alpha} u^{\beta}}{T} Y_{\alpha\beta} 
- \partial_{\beta} \overline{\mu}_{R} \, \nu_{R}^{\beta} - {\mathcal A}_{R}\, \overline{\mu}_{R} Y_{\alpha\beta} \widetilde{Y}^{\alpha\beta}.
\label{th7}
\end{equation}
We remark that the specific definition of the entropy four-vector depends on the chemical potential of the system. However, since the coefficient ${\mathcal A}_{R}$ does not have a definite sign, the 
anomalous currents may even lead to violation of the second principle of thermodynamics 
unless ${\mathcal Z}$ vanishes identically.

\subsection{Magnetic and vortical coefficients}
The vortical and the magnetic currents modify  also 
the diffusive contributions denoted, respectively, by 
 $\nu^{\alpha}$ and $\nu^{\alpha}_{R}$ in Eq. (\ref{th7}). Four different coefficients parametrize the relation between ($\nu^{\alpha}$, $\nu^{\alpha}_{R}$) and 
($\omega^{\alpha}$, ${\mathcal B}^{\alpha}$):
\begin{equation}
\nu^{\alpha} = \Lambda_{\omega} \, \omega^{\alpha} + \Lambda_{B} \, {\mathcal B}^{\alpha}, \qquad 
\nu^{\alpha}_{R} = \Lambda_{R\,\omega} \,\omega^{\alpha} + \Lambda_{R\,B} \, {\mathcal B}^{\alpha},
\label{th8}
\end{equation}
where $(\Lambda_{\omega},\,\Lambda_{B} )$ and $(\Lambda_{R\,\omega}, \, \Lambda_{R\,B})$ all depend on 
the chemical potential and on the temperature. Using Eqs. (\ref{C3}), (\ref{C4}) and (\ref{C5}) the condition  ${\mathcal Z} =0$ together with the explicit expression of ${\mathcal Z}$ (see Eq. (\ref{th7})) becomes:
\begin{eqnarray}
&& \biggl[ 2 {\mathcal S}_{B} - \biggl(\frac{\Lambda_{\omega}}{T}
\biggr)\biggr] \,( \omega^{\alpha} {\mathcal B}_{\alpha})
+ \biggl[4 \overline{\mu}_{R} {\mathcal A}_{R} - \biggl(\frac{\Lambda_{B}}{T}\biggr) \biggr]({\mathcal E}^{\alpha} {\mathcal B}_{\alpha}) 
\nonumber\\
&& - \frac{2}{w} \sigma_{\mathrm{c}}\omega^{\alpha} {\mathcal E}^{\beta}u^{\mu} {\mathcal B}^{\nu}E_{\alpha\beta\mu\nu}{\mathcal S}_{\omega}+ \omega^{\alpha} {\mathcal P}_{\alpha} + {\mathcal B}^{\alpha} {\mathcal Q}_{\alpha} =0 
\label{th9}
\end{eqnarray}
where ${\mathcal P}_{\alpha}$ and ${\mathcal Q}_{\alpha}$ are two differential operators defined respectively, as:
\begin{equation}
{\mathcal P}_{\alpha} = \partial_{\alpha} {\mathcal S}_{\omega} - \frac{2}{w} {\mathcal S}_{\omega} \partial_{\alpha} p - \partial_{\alpha} \overline{\mu}_{R}\,\Lambda_{R\,\omega}, \qquad {\mathcal Q}_{\alpha} = \partial_{\alpha} {\mathcal S}_{B} - \frac{{\mathcal S}_{B}}{w} {\mathcal S}_{\omega} \partial_{\alpha} p - \partial_{\alpha} \overline{\mu}_{R}\, \Lambda_{R\,B}.
\label{th11}
\end{equation}
The results of Eqs. (\ref{th7})--(\ref{th9}) follow easily if we recall that, by definition, $u^{\alpha}\omega_{\alpha}$, 
$u^{\beta} {\mathcal E}_{\beta}$ and $u^{\gamma} {\mathcal B}_{\gamma}$ are all vanishing. 

To satisfy the condition expressed by Eq. (\ref{th9}) the four-vectors multiplying  $\omega^{\alpha}$ and  ${\mathcal B}^{\alpha}$ 
 must vanish together with the coefficients of the terms multiplied by $\omega^{\alpha} {\mathcal B}_{\alpha}$ and ${\mathcal E}^{\alpha} {\mathcal B}_{\alpha}$.  We then arrive at the following conditions:
\begin{equation} 
{\mathcal P}_{\alpha} =0,\qquad  {\mathcal Q}_{\alpha} =0, \qquad \Lambda_{B} = 4 \mu_{R} {\mathcal A}_{R},\qquad \Lambda_{\omega} = 2 T {\mathcal S}_{B}, \qquad {\mathcal S}_{\omega} =0.
\label{th12}
\end{equation}
If, as established, ${\mathcal S}_{\omega}=0$ then Eq. (\ref{th9}) also implies that $\Lambda_{R\,\omega}=0$.
All the coefficients we ought to determine depend on $\overline{\mu}_{R}$ and on the pressure. Thus the conditions of Eq. (\ref{th12}) 
are equivalent to the following system of equations:
\begin{equation}
\biggl( \frac{\partial {\mathcal S}_{B}}{\partial p} - \frac{{\mathcal S}_{B}}{w} \biggr) \partial_{\alpha} p + 
\biggl( \frac{\partial {\mathcal S}_{B}}{\partial \overline{\mu}_{R}} - \Lambda_{R\,B}  \biggr) \partial_{\alpha} \overline{\mu}_{R} =0,
\label{eq1}
\end{equation}
where $\Lambda_{\omega} = 2 T {\mathcal S}_{B}$ and $\Lambda_{B} = 4 {\mathcal A}_{R} \overline{\mu}_{R} T$.
Using some standard thermodynamic relations (giving the partial derivatives of the pressure and of the rescaled chemical 
potential with respect to the temperature) the various kinetic coefficients can be determined, after some algebra:
\begin{eqnarray}
&& {\mathcal S}_{B}(\overline{\mu}_{R}, T) = T \, a_{B}(\overline{\mu}_{R}),\qquad \Lambda_{R\,B} = \frac{\partial}{\partial\overline{\mu}_{R}} \biggl[ T a_{B} (\overline{\mu}_{R})\biggr],
\label{eq5}\\
&& \Lambda_{\omega}(\overline{\mu}_{R}, T) = 2 \,T^2 a_{B}(\overline{\mu}_{R}), \qquad \Lambda_{B}(\overline{\mu}_{R}, T) = 4 \,{\mathcal A}_{R} \, \overline{\mu}_{R}\, T,
\label{eq6}
\end{eqnarray}
where $a_{B}(\overline{\mu}_{R})$ is an arbitrary function of the rescaled chemical potential. Note also that 
$\Lambda_{B}$ is fully determined in terms of the coefficient of the anomaly and it is, in practice, only function 
of the chemical potential itself since, by definition, $\overline{\mu}_{R} T = \mu_{R}$. 

In summary, in a globally neutral plasma with an anomalous current, the relativistic second law implies that the non-anomalous current must contain  magnetic and  vortical contributions. If the plasma is not hypercharge neutral the form of the kinetic coefficients is subjected to a higher degree of arbitrariness  since a second chemical potential must be introduced in the analysis (see the appendix of Ref. \cite{mg2013}). 

\renewcommand{\theequation}{3.\arabic{equation}}
\setcounter{equation}{0}
\section{High-frequency propagation}
\label{sec3}
\subsection{Two-fluid AMHD equations}
The kinetic coefficients of Eqs. (\ref{eq5}) and (\ref{eq6}) can be redefined, for practical reasons, as:
\begin{eqnarray}
c_{\omega}(\overline{\mu}_{R},T) &=& 8 \pi T^2 a_{B}(\overline{\mu}_{R}), \qquad  c_{B}(\overline{\mu}_{R},T) = 16 \pi {\mathcal A}_{R} \overline{\mu}_{R} T,
\nonumber\\
 \qquad c_{RB}(\overline{\mu}_{R},T) &=& \frac{\partial}{\partial\overline{\mu}_{R}} \biggl[ T a_{B} (\overline{\mu}_{R})\biggr], \qquad c_{R\omega}(\overline{\mu}_{R},T)  =0.
\label{Ccoeff}
\end{eqnarray}
The hypermagnetic  and hyperelectric fields denoted, in what follows, by $\vec{B}$ and 
$\vec{E}$ obey the following set of equations 
\begin{eqnarray}
\vec{\nabla} \cdot \vec{E} &=& 4 \pi q (\widetilde{n}_{+} - \widetilde{n}_{-}),\qquad \vec{\nabla} \times \vec{E} = - \partial_{t} \vec{B}, 
\label{S2a}\\
 \vec{\nabla}\times \vec{B} &=& 4\pi q (\widetilde{n}_{+}\, \vec{v}_{+} - 
\widetilde{n}_{-}\, \vec{v}_{-} ) + c_{\omega}\, \vec{\omega} - c_{B} \vec{B} + \partial_{t}\vec{E},
\label{S3}
\end{eqnarray}
where $\vec{B}$ is divergenceless (i.e. $\vec{\nabla} \cdot \vec{B} =0$). 
The coefficients $c_{\omega}(\overline{\mu}_{R},T)$ and $c_{B}(\overline{\mu}_{R},T) $ multiply, respectively, the 
vortical and the magnetic currents of Eq. (\ref{S3}). The three-vector where  $\vec{\omega}$ defines the total vorticity $\vec{\omega} = (\widetilde{\rho}_{+} \vec{\omega}_{+} + \widetilde{\rho}_{-} \vec{\omega}_{-})/(\widetilde{\rho}_{+} + \widetilde{\rho}_{-})$
and should not be confused with the frequency (denoted by $\Omega$ in what follows). The energy densities of the charged species are denoted by $\widetilde{\rho}_{\pm}$. To establish a  direct connection with the spectrum of conventional plasmas in the limit of vanishing kinetic coefficients we shall preferentially consider the situation where 
the charged species are massive, namely 
$\widetilde{\rho}_{\pm} = \widetilde{n}_{\pm} ( m_{\pm} 
+ 3 T_{\pm}/2)$ and $p_{\pm} = \widetilde{n}_{\pm} T_{\pm}$ with $  T_{\pm}/m_{\pm} \ll 1$. 

The remaining coefficients $c_{R\omega}(\overline{\mu}_{R},T)$ and $c_{RB}(\overline{\mu}_{R},T)$ of Eq. (\ref{Ccoeff}) affect directly the evolution of the concentrations:
\begin{eqnarray}
&&\partial_{t} \widetilde{n}_{+} + \vec{\nabla} \cdot( \widetilde{n}_{+} \, \vec{v}_{+}) + \frac{1}{q} \vec{\nabla} \cdot (c_{\omega} \, \lambda_{+} \vec{\omega}_{+} ) -  \frac{1}{q} \vec{\nabla} \cdot (c_{B} \, \lambda_{+} \vec{B}) =0, 
\label{NP}\\
&&\partial_{t} \widetilde{n}_{-} + \vec{\nabla} \cdot( \widetilde{n}_{-} \, \vec{v}_{-}) - \frac{1}{q} \vec{\nabla} \cdot (c_{\omega} \, 
\lambda_{-} \vec{\omega}_{-}) +  \frac{1}{q} \vec{\nabla} \cdot (c_{B} \, \lambda_{-} \vec{B} ) =0,
\label{NM}\\
&&\partial_{t} \widetilde{n}_{R} + \vec{\nabla} \cdot( \widetilde{n}_{R} \, \vec{v}_{R}) + \vec{\nabla} \cdot( c_{R\,\omega} \, \vec{\omega}) -
\vec{\nabla} \cdot( c_{R\,B} \, \vec{B})= - 4 {\mathcal A}_{{\mathcal R}} \vec{E}\cdot\vec{B},
\label{NC}
\end{eqnarray} 
where $\lambda_{\pm} = \widetilde{\rho}_{\pm}/(\widetilde{\rho}_{+} + \widetilde{\rho}_{-} )$.
Concerning Eqs. (\ref{S2a})--(\ref{S3}) and Eqs. (\ref{NP})--(\ref{NC}) few comments are in order. 
 If  $\widetilde{n}_{+} \neq \widetilde{n}_{-}$ a second chemical potential  $\mu_{Y} $ (corresponding to the hypercharge) can be introduced in  Eq. (\ref{th1}). 
 The global hypercharge neutrality of the plasma implies $\mu_{Y}=0$. The peculiar velocities determining the currents 
obey the following set of equations:
\begin{eqnarray}
&& \partial_{t}\vec{v}_{-} + (\vec{v}_{-} \cdot\vec{\nabla})\vec{v}_{-}  = - q\,\frac{ \widetilde{n}_{-}}{\widetilde{\rho}_{-} } [ \vec{E} + \vec{v}_{-} \times \vec{B}] +  \Gamma_{c} ( \vec{v}_{+} - \vec{v}_{-}) - \frac{\vec{\nabla} p_{+}}{\rho_{+}},
\label{SA}\\
&&  \partial_{t} \vec{v}_{+} + (\vec{v}_{+} \cdot\vec{\nabla})\vec{v}_{+} = q\,\frac{\widetilde{n}_{+}}{\widetilde{\rho}_{+} }   [ \vec{E} + \vec{v}_{+} \times \vec{B}] + \Gamma_{c} \frac{\widetilde{\rho}_{-}}{\widetilde{\rho}_{+}}( \vec{v}_{-} - \vec{v}_{+})- \frac{\vec{\nabla} p_{-}}{\rho_{-}},
\label{SB}\\
&& \partial_{t} \vec{v}_{R} + (\vec{v}_{R} \cdot\vec{\nabla})\vec{v}_{R} = 0,
\label{SC}
\end{eqnarray}
where  the pressure gradients shall be eventually neglected; $\Gamma_{c}$ denotes the collision frequency  determining 
 the generalized conductivity in the single fluid limit.

\subsection{Linearization of the two-fluid equations}
Equations 
(\ref{S2a})--(\ref{S3}), (\ref{NP})--(\ref{NC}) and  (\ref{SA})--(\ref{SC}) will now be  linearized in the 
presence of the weak background magnetic field $\vec{B}_{0}$ with the aim of deriving the dispersion relations. The background field 
will be considered homogeneous: this means that the variation of $\vec{B}_{0}$ occurs 
over typical length-scales\footnote{ In section \ref{sec4} we shall specifically discuss 
also the opposite limit where $\vec{B}_{0}$ varies appreciably over typical lengths $L < 1/c_{B}$ 
and we shall see that, in this case, the background solution belongs to the class of Beltrami fields.} much larger than $1/c_{B}$.
The fluctuations of the various quantities will be introduced as follows:
\begin{equation}
\widetilde{n}_{\pm}(t, \vec{x}) = n_{0} + \delta \widetilde{n}_{\pm}( t,\vec{x}),\qquad \widetilde{n}_{R}(t, \vec{x}) = n_{1} + \delta \widetilde{n}_{R}( t,\vec{x}),\qquad \vec{B}(t,\vec{x}) = \vec{B}_{0} + \delta \vec{B}(t,\vec{x}),
\label{fluct}
\end{equation}
while for the other quantities (i.e. $\vec{v}_{\pm}(t,\vec{x}) = 
\delta \vec{v}_{\pm} (t,\vec{x})$, $\vec{v}_{R}(t,\vec{x}) = 
\delta \vec{v}_{R} (t,\vec{x})$ and $\vec{E}(t,\vec{x})= \delta \vec{E}(t,\vec{x})$) the fluctuations coincide with the field itself.
In Eq. (\ref{fluct}) $n_{0}$ and $n_{1}$ are, respectively, the uniform background charge and the uniform chiral concentration.
The homogeneous 
value of the chemical potential is related to $n_{1}$ and the kinetic 
coefficients will also be homogeneous. In the case of approximate thermal 
equilibrium the chemical potential can be related to the concentration as $\overline{\mu}_{R} = \mu_{0} \widetilde{n}_{R}/\varsigma$
where $\varsigma$ denotes the entropy density at equilibrium and where $\mu_{0}$ is a numerical constant. Therefore 
if $\widetilde{n}_{R}$ is perturbed around a homogeneous background the kinetic coefficients 
will also be, in the first approximation homogeneous.  
Thanks to Eq. (\ref{fluct}) the perturbed version of the evolution of the concentrations can be written as:
\begin{equation}
\delta \widetilde{n}_{\pm}' + n_{0} (\vec{\nabla} \cdot \delta \vec{v}_{\pm}) =0,\qquad 
 \delta \widetilde{n}_{R}' + n_{1} (\vec{\nabla} \cdot \delta \vec{v}_{R}) = - 4 {\mathcal A}_{{\mathcal R}} \,\,\delta \vec{E} \cdot \vec{B}_{0},
\label{chiralconc}
\end{equation}
where the prime denotes a derivation with respect to the time coordinate $t$.
Since the kinetic coefficients are homogeneous in the first 
approximation,  their contribution disappears from Eq. (\ref{chiralconc}). 
With the same notations Eqs. (\ref{SA}), (\ref{SB}) and (\ref{SC}) imply instead:
\begin{equation}
\delta\vec{v}_{\pm}^{\,\prime}  = 
\pm \frac{q}{m_{\pm}} \biggl[ \delta \vec{E} 
+ \delta\vec{v_{\pm}}\times \vec{B}_{0}\biggr],\qquad \delta\vec{v}_{R}^{\,\prime} = 0,
\label{deltav}
\end{equation}
where $\Gamma_{c}$ has been neglected but it will become relevant at low frequencies, as we shall see later. 
Finally, after inserting Eq. (\ref{fluct}) into Eqs. (\ref{S2a}) and (\ref{S3}) we obtain:
\begin{eqnarray}
\vec{\nabla} \cdot \delta \vec{E} &=& 4 \pi q ( \delta \widetilde{n}_{+} - \delta \widetilde{n}_{-}),\qquad 
\vec{\nabla} \cdot \delta \vec{B} =0, \qquad \vec{\nabla} \times \delta \vec{E} = -  \delta \vec{B}',
\label{deltacurlE}\\
\vec{\nabla}\times \delta \vec{B} &=&  \delta \vec{E}' + 4\pi\,q\,n_{0}( \delta \vec{v}_{+} -
 \delta \vec{v}_{-})   
 \nonumber\\
 &-& c_{B} \delta \vec{B} + c_{\omega} \biggl[ \lambda_{+} \vec{\nabla}\times \delta \vec{v}_{+} + \lambda_{-} \vec{\nabla}\times \delta \vec{v}_{-} \biggr].
\label{deltacurlB}
\end{eqnarray}
From Eqs. (\ref{deltav})  the equation obeyed by $\delta \vec{\omega}_{\pm}$ can also be deduced and they are 
$\delta \vec{\omega}_{\pm}^{\, \prime} = \pm q [  -  \delta \vec{B}' + \vec{\nabla} \times( \delta \vec{v}_{\pm} \times \vec{B}_{0})]/m_{\pm}$.
Recalling the standard vector identities\footnote{In particular we recall that  $\vec{\nabla} \times (\vec{a} \times \vec{b}) =[\vec{a} ( \vec{\nabla} \cdot\vec{b}) - \vec{b} (\vec{\nabla} \cdot \vec{a}) + (\vec{b}\cdot\vec{\nabla})\vec{a}
- (\vec{a}\cdot\vec{\nabla})\vec{b}$].} the equation for $\delta \vec{\omega}_{\pm}$ can also be expressed as:
\begin{equation}
\delta \vec{\omega}_{\pm}^{\, \prime} = \pm \frac{q}{m_{\pm}}\biggl[  -  \delta \vec{B}' - \vec{B}_{0}  (\vec{\nabla} \cdot \delta \vec{v}_{\pm}) + 
(\vec{B}_{0} \cdot\vec{\nabla})\delta \vec{v}_{\pm}\biggr].
\label{vort1}
\end{equation}
From Eqs. (\ref{deltav}) and (\ref{deltacurlB}) the relevant dispersion relations and the associated refraction indices can be obtained by treating separately the motions parallel and perpendicular to  the magnetic field direction.

\subsection{Appleton-Hartree determinant}
While in conventional plasmas the Appleton-Hartree dispersion relation has been  extensively discussed in the literature \cite{stix,wop}, the AMHD equations 
linearized in the two-fluid limit contain vortical and magnetic currents. The Laplace transform of Eq. (\ref{deltacurlB}) implies the following equation
\begin{equation}
(\vec{\nabla} \times \delta\vec{B})_{\Omega} = - i \,\Omega \,\varepsilon_{s}(\Omega)\cdot \delta \vec{E}_{\Omega}
- c_{B} \, \delta \vec{B}_{\Omega} + i\, c_{\omega} \vec{\nabla} \times [ \varepsilon_{v}(\Omega)\cdot \delta \vec{E}_{\Omega}],
\label{deltacurlB2}
\end{equation}
where $\Omega$ is the frequency (not to be confused with the total vorticity). In Eq. (\ref{deltacurlB2}) $\varepsilon_{s}(\Omega)$ and $\varepsilon_{v}(\Omega)$ denote, respectively, the standard and the vortical components 
of the dielectric tensor. The explicit form of $\varepsilon_{s}(\Omega)$ and $\varepsilon_{v}(\Omega)$ can be found 
in appendix \ref{APPA}; taking then the curl 
of Eq. (\ref{deltacurlE}) and using Eq. (\ref{deltacurlB2}) we obtain the following equation:
\begin{equation}
\vec{\nabla}\times (\vec{\nabla} \times\delta \vec{E}_{\Omega}) = \Omega^2 \varepsilon_{s}(\Omega) \cdot \delta \vec{E}_{\Omega} - c_{B} \vec{\nabla}\times \delta \vec{E}_{\Omega} - \Omega\, c_{\omega}\, \vec{\nabla}\times [ \varepsilon_{v}(\Omega) \cdot \delta \vec{E}_{\Omega}].
\label{disp1}
\end{equation}
We can now go to Fourier space and write Eq. (\ref{disp1}) as:
\begin{equation}
- \vec{k} \times \vec{k} \times \delta \vec{E}_{\vec{k}\, \Omega} = \Omega^2 \,\varepsilon_{s}(\Omega)\cdot \delta \vec{E}_{\vec{k}\, \Omega}
- i\, c_{B} \, \vec{k} \times \delta \vec{E}_{\vec{k}\, \Omega} - i\, c_{\omega} \, \Omega\,\vec{k}\times [ \varepsilon_{v}(\Omega) \cdot \delta \vec{E}_{\vec{k}\, \Omega}].
\label{disp2}
\end{equation} 
We can therefore introduce the refractive index\footnote{The refractive index cannot be confused 
with the concentrations denoted by $\widetilde{n}_{\pm}$ and $\widetilde{n}_{R}$; their homogeneous
values $n_{0}$ and $n_{1}$ carry specific subscripts so that the notations are clearly established.}
$n$ satisfying $ n = k/\Omega$ where $k= |\vec{k}|$;
choosing the coordinate system as $\vec{k} =(0,\,\, n\Omega \sin{\theta}, \,\, n \Omega \cos{\theta})$ we can obtain 
from Eq. (\ref{disp2}) the following Appleton-Hartree matrix:
\begin{equation}
 \left(\matrix{\bigl[1 - \frac{\varepsilon_{1}}{n^2} + c_{\omega} \frac{\varepsilon_{4}}{n} c(\theta)\bigr]
& - i\bigl[\frac{\varepsilon_{2}}{n^2} 
+ \frac{c_{B}}{n \Omega} c(\theta) +\frac{c_{\omega}}{n } \varepsilon_{3} c(\theta) \bigr] & i \frac{c_{B}}{n \Omega}\, s(\theta) &\cr
i\bigl[\frac{\varepsilon_{2}}{n^2} 
+ \frac{c_{B}}{n \omega} c(\theta)+\frac{c_{\omega}}{n } \varepsilon_{3} c(\theta) \bigr] 
& \bigl[ c^2(\theta) - \frac{\varepsilon_1}{n^2} + \frac{\varepsilon_{4}(\Omega)}{n} c_{\omega} c(\theta) \bigr] 
& - s(\theta) c(\theta)&\cr
- i \frac{c_{B}}{n \Omega} s(\theta) - i \frac{c_{\omega}}{n} \varepsilon_{3}(\Omega) & - s(\theta)c(\theta) + \frac{c_{\omega}}{n} \varepsilon_{4} s(\theta)
&\bigl[ s^2(\theta) -\frac{\varepsilon_{\parallel}(\omega)}{n^2}\bigr] }\right).
\label{matrix}
\end{equation}
The above matrix reduces to the standard form of the Appleton-Hartree matrix in the limit $c_{\omega} \to 0$ and $c_{B} \to 0$. 

\renewcommand{\theequation}{4.\arabic{equation}}
\setcounter{equation}{0}
\section{Dispersion relations}
\label{sec4}
The determinant of the Appleton-Hartree matrix obtained in Eq. (\ref{matrix}) leads to the following expression:
\begin{eqnarray}
&& \sin^2{\theta}(\varepsilon_{\parallel} - n^2) [n^2 ( \varepsilon_{L} + \varepsilon_{R}) - 2 \varepsilon_R \, \varepsilon_{L}] - 
2  \cos^2{\theta}\varepsilon_{\parallel}( n^2 - \varepsilon_{L}) (n^2 - \varepsilon_{R}) 
\nonumber\\
&&+ 2n^6 \bigl[ c_{B}^2 f_{B}(\varepsilon,\, \Omega,\, n,\, \theta) + c_{B}  g_{B}(\varepsilon,\, \Omega,\, n,\, \theta) + c_{\omega}^2 f_{\omega}(\varepsilon,\, \Omega,\, n,\, \theta) + c_{\omega}  g_{\omega}(\varepsilon,\, \Omega,\, n,\, \theta) 
\nonumber\\
&& c_{B}^2\, c_{\omega} \, h_{1} (\varepsilon,\, \Omega,\, n,\, \theta) + c_{B} c_{\omega}  \, h_{2} (\varepsilon,\, \Omega,\, n,\, \theta)
+ c_{\omega}^2 \, c_{B}  \, h_{3} (\varepsilon,\, \Omega,\, n,\, \theta)\bigr] =0.
\label{AH}
\end{eqnarray}
Equation (\ref{AH}) is written in terms of the $7$ functions explicitly reported in Eq. (\ref{seven}) of appendix \ref{APPA}.  These functions have a specific dependence upon the dielectric tensors; with a collective notation such a dependence has been indicated by $\varepsilon$.  
The notations followed in Eq. (\ref{AH}) imply that $c_{B}^2$ multiplies $f_{B}$, $c_{\omega}^2$ multiplies $f_{\omega}$; $g_{B}$ and $g_{\omega}$ multiply, respectively, $c_{B}$ and $c_{\omega}$; the three functions $h_{1}$, $h_{2}$ and $h_{3}$ multiply 
instead the mixed products. 
Finally both in Eqs.
(\ref{AH}) and in Eq. (\ref{seven}) we have introduced $\varepsilon_{L}= (\varepsilon_{1}+\varepsilon_{2})$ and $\varepsilon_{R}=  (\varepsilon_{1}-\varepsilon_{2})$
given by:
\begin{eqnarray}
\varepsilon_{L}(\Omega) = 1 - \frac{\Omega_{p}^2}{(\Omega + \Omega_{B\,-}) (\Omega - \Omega_{B\,+})}, \qquad 
\varepsilon_{R}(\Omega) = 1 - \frac{\Omega_{p}^2}{(\Omega + \Omega_{B\,+}) (\Omega - \Omega_{B\,-})},
\label{epsLR}
\end{eqnarray}
where $\Omega_{p}^2 = (\Omega_{p\,+}^2 + \Omega_{p\, -}^2)$.
When $c_{B} = c_{\omega} =0$ the magnetic and the vortical currents are absent 
from the two-fluid AMHD equations and Eq. (\ref{AH}) implies the standard result \cite{stix}:
\begin{equation}
\sin^2{\theta} \biggl(\frac{1}{n^2} - \frac{1}{\varepsilon_{\parallel}}\biggr) \biggl[ \frac{1}{2} \biggl(\frac{1}{\varepsilon_{L}} + \frac{1}{\varepsilon_{R}}\biggr) - \frac{1}{n^2}\biggr] = \cos^2{\theta} \biggl(\frac{1}{\varepsilon_{L}} - \frac{1}{n^2}\biggr) \biggl(\frac{1}{\varepsilon_{R}} - \frac{1}{n^2}\biggr).
\label{AH2}
\end{equation}
The dispersion relations for a wave propagating parallel (i.e. $ \theta =0$) and 
perpendicular (i.e. $\theta = \pi/2$) to the magnetic field direction can be obtained easily derived from Eq. (\ref{AH2}). If $\theta=0$  Eq. (\ref{AH2}) reduces to $(n^2 - \varepsilon_{\rm R})(n^2 -\varepsilon_{\rm L})=0$
while for $\theta=\pi/2$ Eq. (\ref{AH2}) implies
 $(n^2 - \varepsilon_{\parallel}) [ n^2 (\varepsilon_{\rm L} + \varepsilon_{\rm R})
- 2 \varepsilon_{\rm L} \varepsilon_{\rm R} ] =0$. These dispersion relations 
give therefore the conventional results\footnote{Along $\theta=0$ we thus obtain usual dispersion relations for the two circular 
polarizations of the electromagnetic wave, i.e. $n^2 = \varepsilon_{\rm R}$ 
and $n^2 = \varepsilon_{\rm L}$, while along $\theta=\pi/2$
we have the dispersion relations for the ``ordinary'' 
(i.e. $n^2 = \varepsilon_{\parallel}$) and ``extraordinary'' (i.e. $n^2 = 
2 \varepsilon_{\rm R} \varepsilon_{\rm L}/(\varepsilon_{\rm R} + \varepsilon_{\rm L})$)
plasma waves.} which will be generalized hereunder.

\subsection{Free-field propagation}
In the absence of magnetic field there is no preferred direction and the dispersion relations follow from Eqs. (\ref{AH2}) and (\ref{seven}) by setting all the Larmor frequencies to zero. In this case
$\varepsilon_{R} = \varepsilon_{L} = \varepsilon_{\parallel}$ and the dispersion 
relations stem from the following two conditions, namely:
\begin{equation}
\varepsilon_{\parallel}(\Omega)=0,\qquad (n^2 - \varepsilon_{\parallel}) \Omega \mp n c_{B} =0.
\label{FF1}
\end{equation}
Equation (\ref{FF1}) demonstrates that the vortical current does not contribute to the dispersion relations in the free-field case:  $c_{\omega}$ is absent from Eq. (\ref{FF1}) since the two-fluid effects cancel in the total vorticity. This cancellation is either exact (as in the case of the free-field propagation) or approximate (as we shall see later 
in the presence of the magnetic field).  Indeed, as it can be explicitly verified from Eqs. (\ref{epstenSV}), (\ref{eps3}) and (\ref{eps4}),  $\epsilon_{v}(\Omega) \to 0$ when $B_{0} \to 0$: in the limit $B_{0} \to 0$ the vorticity of positively and negatively charged species is balanced so that the net total vorticity vanishes.

The dispersion relation $\varepsilon_{\parallel}(\Omega)=0$ implies 
$\Omega = \Omega_{p}$.  This wave does not propagate since its group 
velocity vanishes and these are nothing but the electrostatic plasma oscillations \cite{stix}.
The solution of the second equation in Eq. (\ref{FF1}) is
instead\footnote{The positive square root has been chosen in Eq. (\ref{FF2}) in order to get $\Omega >0$; we consider only positive $\Omega$ since solutions with $\Omega < 0$ simply correspond to waves travelling in the opposite direction.} 
\begin{equation}
n= \pm \frac{c_{B}}{2 \Omega} + \sqrt{ 1 - \frac{\Omega_{p}^2}{\Omega^2} + \frac{c^2_{B}}{4 \Omega^2}}.
\label{FF2}
\end{equation}
 Equation (\ref{FF2}) implies also $\Omega^2 = \Omega_{p}^2 + k^2 \mp k c_{B}$;  these modes are propagating but only affected by the magnetic current, as previously remarked. The birefringent nature 
 of the dispersion relations will be discussed a bit later since this free-field effect may interfere with 
 the presence of the background magnetic field. 
 
If $c_{B} \to 0$ and $c_{\omega}\to 0$ we have that $n\to 0$ whenever one of the following three possibility are separately verified
$ \varepsilon_{\parallel}(\Omega) =0$ or  $\varepsilon_{L}(\Omega) =0$ or even $\varepsilon_{R}(\Omega) =0$.
The frequencies arising from the previous conditions are cut-offs because, for given equilibrium conditions, they define frequencies above or below which the wave ceases to propagate at any angle ($k \to 0$ for finite $\Omega$, i.e. $v_{p} = \Omega/k\to \infty$). This is what happens, in particular, with the dispersion relation of Eq. (\ref{FF2}). Let us finally 
remark that the remaining two cut-offs stemming from the conditions $\varepsilon_{L}(\Omega) =0$ and $\varepsilon_{R}(\Omega) =0$ in 
Eqs. (\ref{epsLR}) are given, respectivey, by:
\begin{eqnarray}
\Omega_{R} &=& \sqrt{\Omega_{p}^2 + (\Omega_{B\,+} + \Omega_{B\,-})^2/4} - (\Omega_{B\,+} + \Omega_{B\,-})/2,
\label{FF5a}\\
\Omega_{L} &=& \sqrt{\Omega_{p}^2 + (\Omega_{B\,+} + \Omega_{B\,-})^2/4} + (\Omega_{B\,+} + \Omega_{B\,-})/2.
\label{FF5b}
\end{eqnarray}
 
\subsection{Parallel propagation}

Taking the limit $\theta \to 0$ in Eq. (\ref{AH}) and recalling the results of Eq. (\ref{seven}) we obtain:
\begin{equation}
\varepsilon_{\parallel} \{ n\,c_{B}  + [n^2 + n\,c_{\omega}\,(\varepsilon_{3} + \varepsilon_{4}) - \varepsilon_{R}]\Omega\}
\{n\,c_{B}  - [n^2   + n \,c_{\omega} \, (-\varepsilon_{3}  + \varepsilon_{4}) -\varepsilon_{L}] \Omega\}=0.
 \label{AH3}
\end{equation}   
If  $\varepsilon_{\parallel}(\Omega)= 0$ we go back to the case of electrostatic oscillations.
Therefore, assuming $\varepsilon_{\parallel}(\Omega)\neq 0$, Eq. (\ref{AH3}) implies that the standard
dispersion relations are modified as: 
\begin{eqnarray}
n &=& \frac{1}{2 \Omega} \biggl[ - c_{B} - c_{\omega} ( \varepsilon_{3} +\varepsilon_{4} )\Omega \pm \sqrt{ 4 \varepsilon_{R} \Omega^2 +
[ c_{B} + c_{\omega} \Omega (\varepsilon_{3} + \varepsilon_{4}) ]^2}\biggr],
\label{AH3a}\\
n &=&\frac{1}{2 \Omega}  \biggl[  c_{B} + c_{\omega} ( \varepsilon_{3} -\varepsilon_{4} )\Omega \pm \sqrt{ 4 \varepsilon_{L} \Omega^2 +
[ c_{B} + c_{\omega} \Omega (\varepsilon_{3} - \varepsilon_{4}) ]^2}\biggr].
\label{AH3b}
\end{eqnarray}
Thus the dispersion relations for the generalized $L$-mode and $R$-mode are given, respectively, by:
\begin{eqnarray}
&& \Omega^2\, \varepsilon_{R}(\Omega) = k^2 + k [ c_{B} + c_{\omega} (\varepsilon_{3}+\varepsilon_{4}) \Omega],
\label{Rmode}\\
&& \Omega^2\, \varepsilon_{L}(\Omega) = k^2 - k [ c_{B} + c_{\omega} (\varepsilon_{3}-\varepsilon_{4}) \Omega].
\label{Lmode}
\end{eqnarray}
In the high-frequency limit (i.e. formally  $\Omega \to \infty$) we have that $c_{\omega} (\varepsilon_{3}\pm\varepsilon_{4}) \Omega \to 0$ since,  from Eqs. (\ref{eps3})--(\ref{eps4}),  we have:
\begin{eqnarray}
&& c_{\omega}(\varepsilon_{3} + \varepsilon_{4}) \Omega = \frac{q \, c_{\omega}}{(m_{+}+m_{-})}\biggl[ \frac{\Omega}{\Omega_{B\, -} - \Omega} + \frac{\Omega}{\Omega_{B\,+} + \Omega}\biggr],
\label{lim1}\\
&& c_{\omega}(\varepsilon_{3} - \varepsilon_{4}) \Omega = \frac{q \, c_{\omega}}{(m_{+} + m_{-})}\biggl[ \frac{\Omega}{\Omega- \Omega_{B\, +} } - \frac{\Omega}{\Omega + \Omega_{B\,-}}\biggr].
\label{lim2}
\end{eqnarray}

The results of Eqs. (\ref{Rmode})--(\ref{Lmode}) and (\ref{lim1})--(\ref{lim2}) demonstrate, once more,  that in the high-frequency limit of the spectrum the magnetic current dominates against the vortical current.
For intermediate frequencies  (i.e. as soon as we reduce $\Omega$) the terms containing the natural frequencies of the plasma come then into play so that for the $R$ and $L$ modes the corresponding dispersion 
relations become:
\begin{eqnarray}
&& \Omega^2 = k^2 + k c_{B} + \frac{\Omega_{p}^2 \Omega}{(\Omega- \Omega_{B\,-})}, \qquad R-\mathrm{ mode},
\nonumber\\
&& \Omega^2 = k^2 - k c_{B} + \frac{\Omega_{p}^2 \Omega}{(\Omega+ \Omega_{B\,-})}, \qquad L-\mathrm{ mode}.
\label{lim3}
\end{eqnarray}
As in the standard case, the phase velocity of the $R$-mode is greater than that of the $L$-mode. In Eq. (\ref{lim3}) we assumed $m_{+} > m_{-}$ and therefore 
$\Omega_{B\,+}< \Omega_{B\,-}$. In the limit $k\to 0$ the  $R$-mode cut-off occurs above $\Omega_{p}$ while 
the $L$-mode cut-off occurs below $\Omega_{p}$ (i.e., recalling Eqs. (\ref{FF5a}) and (\ref{FF5b}),  $\Omega\to \Omega_{R}$ and $\Omega\to \Omega_{L}$). In the low-frequency limit $\varepsilon_{R}$ and $\varepsilon_{L}$ coincide to 
leading order in $(\Omega/\Omega_{B\,+})$ and in  $(\Omega/\Omega_{B\,-})$ since 
\begin{equation}
\lim_{\Omega\to 0} \varepsilon_{R}(\Omega) = \lim_{\Omega\to 0} \varepsilon_{L}(\Omega) \to 1 + \frac{\Omega_{p}^2}{\Omega_{B\,+} \, \Omega_{B\,-}} =1 + 
\frac{1}{v_{A}^2},\qquad v_{A} = \frac{B_{0}}{\sqrt{4 \pi n_{0} (m_{+} + m_{-})}},
\label{lim4}
\end{equation}
where $v_{A}$ denotes the Alfv\'en velocity of the system. In the low-frequency limit the dispersion relations for the $R$-mode and for the $L$-mode 
are, respectively,
\begin{eqnarray}
&& \Omega^2 = \frac{v_{A}^2}{1 + v_{A}^2}\biggl\{ k^2 + k \biggl[c_{B} + \frac{q\, c_{\omega}}{m}\biggl(\frac{\Omega}{\Omega_{B\,+}} + \frac{\Omega}{\Omega_{B\,-}}  \biggr) \biggr]\biggr\},
\label{lim5}\\
&& \Omega^2 = \frac{v_{A}^2}{1 + v_{A}^2}\biggl\{ k^2 - k \biggl[c_{B} - \frac{q\,c_{\omega}}{m}\biggl(\frac{\Omega}{\Omega_{B\,+}} + \frac{\Omega}{\Omega_{B\,-}}  \biggr)  \biggr]\biggr\},
\label{lim6}
\end{eqnarray}
since $v_{A} \ll 1$ we also have $v_{A}^2/(1 + v_{A}^2)\simeq v_{A}^2$.

Having determined the dispersion relations in the case of parallel propagation,  the Faraday rotation 
rate can be easily determined with the standard procedure. 
The generalized Faraday rotation angle experienced 
by the linearly polarized radiation travelling parallel to the magnetic field direction can be obtained as 
\begin{equation}
\Delta\Phi = \frac{\Omega}{2} \biggl\{ \frac{c_{B}}{\Omega} + c_{\omega} \varepsilon_{3}
+ \sqrt{\varepsilon_{L} + \biggl[\frac{c_{B}}{2\Omega} + \frac{c_{\omega}}{2} (\varepsilon_{3} + \varepsilon_{4})\biggr]^2}
- \sqrt{\varepsilon_{R} + \biggl[\frac{c_{B}}{2\Omega} + \frac{c_{\omega}}{2} (\varepsilon_{3} - \varepsilon_{4})\biggr]^2}\biggr\}  \Delta {\rm L},
\label{FR}
\end{equation}
where $\Delta {\rm L}$ is 
the distance travelled by the signal in the direction parallel to the 
magnetic field direction. It is interesting to compare the contribution of the terms depending 
upon $c_{B}$ and those depending upon the background magnetic 
field intensity, i.e. the terms appearing in the squared brackets.
Recalling the expressions of $(\varepsilon_{R},\, \varepsilon_{L})$ we have that 
$\Omega_{B\,+} \ll \Omega_{B\,-}$ and  $\Omega_{p\,+} \ll \Omega_{p\,-}$ (always assuming $m_{+}\gg m_{-}$).
In this case $\Delta \Phi/\Delta L $ interpolates between the standard result $(\Omega_{B\,-}/2)
(\Omega_{p\,-}/\Omega)^2$ (valid when $c_{B} \to 0$) and the constant rotation rate $c_{B}/2$ (valid when 
$B_{0} \to 0$ as in the case of free-field propagation). As it can be explicitly verified the $c_{\omega}$ is  subdominant at high frequencies and can be neglected.

\subsection{Orthogonal propagation}
By setting $\theta \to \pi/2$ in Eq. (\ref{AH}) and recalling the results 
of Eq. (\ref{seven}) we obtain the following simple equation:
\begin{eqnarray}
&& n^4 (\varepsilon_{L} + \varepsilon_{R}) \Omega^2 - n^2 \{c_{B}^2 (\varepsilon_{L} + \varepsilon_{R}) +
c_{B} c_{\omega} [\varepsilon_{4} (-\varepsilon_{L} + \varepsilon_{R}) +\varepsilon_{3} (\varepsilon_{L}  + 
\varepsilon_{R})] \Omega
\nonumber\\
&&+[ \varepsilon_{\parallel} \varepsilon_{R}  + \varepsilon_{L} (\varepsilon_{\parallel} +2\varepsilon_{R})] \Omega^2\} + 2 \varepsilon_{L} \varepsilon_{R} \Omega^2 \varepsilon_{\parallel} =0.
\label{AH4}
\end{eqnarray}
The solution of Eq. (\ref{AH4}) can be obtained by first solving in terms of $n^2$. The result is 
\begin{eqnarray}
n^2 &=& \frac{{\mathcal J}(\varepsilon,\, \Omega) \pm \sqrt{{\mathcal M}(\varepsilon,\, \Omega)}}{2 (\varepsilon_{R} + \varepsilon_{L})\,\Omega^2},
\label{AH4a}\\
{\mathcal J}(\varepsilon,\, \Omega) &=&c_{B}^2 (\varepsilon_{L} +\varepsilon_{R}) + 
 c_{B} \, c_{\omega}\, [\varepsilon_{4} (-\varepsilon_{L} + \varepsilon_{R}) +\varepsilon_{3}
(\varepsilon_{L} +\varepsilon_{R})] \Omega
\nonumber\\
 &+ &  [\varepsilon_{\parallel}\varepsilon_{R}  +  \varepsilon_{L} (\varepsilon_{\parallel}+ 2 \varepsilon_{R})] \Omega^2,
\label{JJ1}\\
{\mathcal M}(\varepsilon,\, \Omega) &=& -8 \varepsilon_{\parallel}\varepsilon_{R}\varepsilon_{L} (\varepsilon_{L} +\varepsilon_{R})
\Omega^4 + \{c_{B}^2 (\varepsilon_{L} +\varepsilon_{R})
\nonumber\\
&+& c_{B} c_{\omega} [\varepsilon_{4} (\varepsilon_{R} - \varepsilon_{L}) + 
\varepsilon_{3} (\varepsilon_{L} +\varepsilon_{R}) ] \Omega +
[ \varepsilon_{\parallel}\varepsilon_{R} + \varepsilon_{L} (\varepsilon_{\parallel} +  2 \varepsilon_{R})] \Omega^2\}^2.
\label{MM1}         
\end{eqnarray}
Equation (\ref{AH4a}) in the limit $c_{\omega} \to 0$ and $c_{B} \to 0$ reduces to the ordinary mode if we choose the plus (i.e.  $n^2 = \varepsilon_{\parallel}$) and to the extraordinary mode (i.e. $n^2 = 
2 \varepsilon_{\rm R} \varepsilon_{\rm L}/(\varepsilon_{\rm R} + \varepsilon_{\rm L})$)
if we choose the minus. In the high-frequency limit the terms multiplying the vortical current are always negligible as already remarked in the case of the parallel propagation. The phenomena related to the oblique propagation will not be specifically discussed.

\subsection{Spectrum around a hypermagnetic knot}
Introducing the three mutually orthogonal unit vectors $\hat{a}(z,p)$, $\hat{b}(z,p)$ and $\hat{z}$ defined in appendix \ref{APPB},
we can consider the modes of fluctuation of the hypermagnetic field around a fully inhomogeneous background $\vec{B}_{0}(t, \vec{x})$, namely: 
\begin{equation}
\vec{B}(t,\vec{x}) = \vec{B}_{0}(t, \vec{x}) + \delta \vec{B}(t, \vec{x}).
\end{equation}
Since the background solution is not uniform we can 
align $\vec{B}_{0}$ along $\hat{a}(z,p)$ and write that $\vec{B}_{0}(z) = B_{0} \, \hat{a}(z,p)$. 
The background equations are solved by setting $p = - c_{B}$ (since $\vec{\nabla}\times \vec{B}_{0} = p \, \vec{B}_{0}$). 
As in the homogeneous case the velocites vanish on the background solution.  For $L < 1/c_{B}$ the background field is homogeneous and the previous analyses 
apply. For typical length-scales larger than the scale of spatial variation of hypermagnetic knot (i.e.  $L \gg 1/c_{B}$)  
there are two separate possibilities for the perturbed velocity field:
either $\delta\vec{v} \parallel \vec{B}_{0}$ 
or $\delta\vec{v} \perp \vec{B}_{0}$.  These two cases will now be separately examined.

The case of parallel propagation mirrors exactly the one already 
discussed in the case of uniform field. If we assume that $\delta \vec{v} \parallel \vec{B}_{0}$ 
 the dispersion relations follow from
\begin{equation}
\{ [k^2 - \Omega^2 \varepsilon_{\parallel}(\Omega)]^2 - c_{B}^2 k^2\}\varepsilon_{\parallel}(\Omega)=0.
\end{equation}
The parallel dielectric tensor is $\varepsilon_{\parallel}(\Omega) = 1 - \Omega_{p}^2/[\Omega (\Omega + i \Gamma_{c})]$  where the 
correction coming from the collision rate has been added for immediate convenience.
The solution $\varepsilon_{\parallel}(\Omega) =0$ gives, as before, the electrostatic wave.  
The solution of $\{ [k^2 - \Omega^2 \varepsilon_{\parallel}(\Omega)]^2 - c_{B}^2 k^2\}=0$ gives, respectively, a high-frequency  and a low-frequency branch.
The high-frequency branch has the same dispersion relation of the free-field case, namely $\Omega^2 \simeq \Omega_{p}^2 + k^2 \mp k c_{B}$. The low-frequency branch is instead derived from the explicit form of the dispersion relation written as:
\begin{equation}
\Omega + i \Gamma_{c} = \frac{\Omega^2}{k^2} ( \Omega + i \Gamma_{c}) - \frac{\Omega_{p}^2\Omega}{k^2} \pm \frac{c_{B}}{k}( \Omega + i \Gamma_{c});
\end{equation}
neglecting the first term at the right-hand side of the previous equation (which is unimportant at low frequencies) we have that:
\begin{equation}
\Omega = - \frac{i \Gamma_{c}(1 \mp c_{B}/k)}{1 + \Omega_{p}^2/k^2 \mp c_{B}/k}.
\end{equation}
The low-frequency mode, in which the conducting current dominates over the 
displacement current, has no counterpart in vacuum. In the low-frequency mode, a small 
electric field proportional to $\Gamma_{c}$ exist to give the necessary current parallel 
to the magnetic field. In the limit $\Gamma_{c} \to 0$ both the electric field and resistivity 
vanish and the low-frequency mode becomes the force-free field. As expected the same phenomenon 
occurs in the absence of magnetic and vortical currents \cite{chu,saling}. 

In the case of orthogonal propagation the fluctuations of the hypermagnetic field and of the velocity can be
decomposed by using the gyrotropic basis of appendix \ref{APPB}:
\begin{equation}
\delta\vec{B}(t, z)  = \delta B_{1}(t) \hat{b}(z,p) + \delta B_{2}(t) \hat{z}, \qquad 
\delta\vec{v}^{(\pm)}(t, z) = \delta v^{(\pm)}_{1}(t) \hat{b}(z,p) + \delta v^{(\pm)}_{2}(t) \hat{z}.
\end{equation}
For a generic velocity fluctuation orthogonal to $\vec{B}_{0}$ we have $\delta \vec{v} \times \vec{B}_{0} = [B_{0} (\delta \vec{v} \cdot \hat{z}) \hat{b} - B_{0}  (\delta \vec{v} \cdot \hat{b})\hat{z}]$; the solutions for $ \delta v^{(\pm)}_{1}(t)$ and $\delta v^{(\pm)}_{2}(t)$ can then be 
expressed as:
\begin{eqnarray}
\delta v_{1}^{(\pm)}(\Omega)&=& \frac{q}{m_{\pm}(\Omega_{B\, \pm}^2 -\Omega^2)} \biggl[ \pm i \Omega \,\delta E_{1} +\Omega_{B\,\pm} \,\delta E_{2} \biggr],
\nonumber\\
\delta v_{2}^{(\pm)}(\Omega) &=&    \frac{q}{m_{\pm}(\Omega_{B\, \pm}^2 -\Omega^2)} \biggl[ \pm i \Omega \,\delta E_{2} -  \Omega_{B\,\pm} \,\delta E_{1} \biggr].
\end{eqnarray}
The dispersion relations  in this case are given by
\begin{equation}
\varepsilon_{R} \varepsilon_{L} - \frac{c_{B} \, c_{\omega}}{2 \,\Omega}\biggl[ \varepsilon_{L} ( \varepsilon_{3} - \varepsilon_{4}) + \varepsilon_{R}  ( \varepsilon_{3} + \varepsilon_{4})\biggr] =0.
\label{disp2a}
\end{equation} 
 In the high-frequency limits defined by Eqs. (\ref{lim1})--(\ref{lim2}), Eq. (\ref{disp2a}) is satisfied 
if $\varepsilon_{R} \varepsilon_{L} = 0$ which is verified when either $\varepsilon_{L}$ or $\varepsilon_{R}$ are vanishing. 
Equation (\ref{disp2a}) leads to vanishing group velocity in the high-frequency regime: the corresponding modes are then not propagating. The 
proper frequencies defined by these equations have been already derived in Eqs. (\ref{Rmode}) and (\ref{Lmode}). 
\renewcommand{\theequation}{5.\arabic{equation}}
\setcounter{equation}{0}
\section{Single fluid description and its implications}
\label{sec5}
The two-fluid equations can now be combined with the purpose of deriving the effective single fluid description 
valid for sufficiently large length-scales and for frequencies much smaller than $\Omega_{p}$ and $\Omega_{B\,\pm}$.  The one-fluid variables are the total current $\vec{J} = q \, (n_{+},\, \vec{v}_{+} - n_{-} \vec{v}_{-})$, the bulk velocity 
of the plasma $\vec{v}=(m_{+} \vec{v}_{+} + m_{-} \vec{v}_{-})/(m_{+} + m_{-})$ and the total mass density $\rho_{m}=(m_{+} n_{+} + m_{-} n_{-})$.
In the globally neutral case $\vec{J}$ and $\rho_{m}$ become, respectively, $\vec{J} = n_{0} (\vec{v}_{+} - \vec{v}_{-})$ and $\rho_{m} = n_{0} (m_{+} + m_{-})$. Summing-up Eq. (\ref{SA}) (multiplied by $m_{+}$) and Eq. (\ref{SB}) (multiplied by $m_{-}$) the evolution 
equation for the bulk velocity of the plasma is  
\begin{equation}
\rho_{m} \biggl[ \partial_{t} \vec{v} + \vec{v}\cdot \vec{\nabla} \vec{v}\biggr] = \vec{J} \times \vec{B} - \vec{\nabla} P + \eta \nabla^2 \vec{v},
\label{onef2}
\end{equation}
where the shear viscosity contribution, labeled by $\eta$, has been added for convenience\footnote{If the total pressure does not vanish 
Eq. (\ref{onef2}) is modified as follows $\partial_{t} [ w \,\vec{v}] + (\vec{v} \cdot \vec{\nabla}) [ w\, \vec{v}]  + \vec{v}\, \vec{\nabla}\cdot[ w \vec{v}] = - \vec{\nabla}P + 
\vec{J} \times \vec{B} + \eta  \nabla^2 \vec{v}$
where $w$, as already discussed, is the enthalpy density.}.  Taking the difference of Eqs. (\ref{SA})  and (\ref{SB}) (and assuming $m_{+} > m_{-}$)  the generalized Ohm's law can be written as:
\begin{equation}
\partial_{t} \vec{J} + \Gamma_{c} \vec{J} \simeq \frac{\Omega_{P\,-}^2}{4\pi}\biggl(\vec{E} + \vec{v}\times \vec{B} + \frac{\vec{\nabla} p_{-}}{q n_{0}} - 
\frac{\vec{J}\times \vec{B}}{q n_{0}} \biggr),
\label{onef3}
\end{equation}
where global neutrality has been assumed. Note that we have also kept the thermoelectric term (depending on the pressure 
gradient of the lightest charge carriers) and the Hall term. Since we shall mainly consider the case of homogeneous pressures 
the thermoelectric term will be neglected; the Hall term is a higher order contribution, as we shall argue. 

In the globally neutral case the single fluid equations stipulate that $\vec{E}$, $\vec{B}$ and $\vec{J}$ are all solenoidal (i.e.
$\vec{\nabla}\cdot \vec{E} =\vec{\nabla}\cdot \vec{B} =\vec{\nabla} \cdot\vec{J}=0$).
A fourth possible solenoidal vector is the bulk velocity of the plasma $\vec{v}$. Indeed,
since the evolution of $\rho_{m}$ and $\rho_{q} =q(n_{+} - n_{-})$ is given by\footnote{ Note that the global neutrality implies $\rho_{q} =0$ and Eq. (\ref{cc3})
demands $\vec{\nabla}\cdot\vec{J}=0$ in full agreement with the solenoidal nature of the total Ohmic current.}
\begin{equation}
\partial_{t} \rho_{m} + \vec{\nabla}\cdot (\rho_{m} \vec{v}) =0, \qquad \partial_{t} \rho_{q}+ \vec{\nabla}\cdot \vec{J}=0,
\label{cc3}
\end{equation}
the incompressible closure $\vec{\nabla}\cdot \vec{v} =0$ will be adopted; consistently with the incompressible closure 
$\rho_{m}$ will be considered homogeneous, at least in the first part of this section. 
A full discussion of other possible closure (such as the ones 
conventionally adopted in conventional plasmas) is desirable but beyond the scoped of this analysis.
The remaining one-fluid equation containing the vortical and the magnetic currents, can be written as:
\begin{equation}
 \vec{\nabla}\times \vec{B} - \partial_{t}\vec{E}= 
4\pi \vec{J} + c_{\omega} \vec{\omega} - c_{B} \vec{B}, \qquad \vec{\omega} = \vec{\nabla} \times \vec{v}.
\label{MX2}
\end{equation}
Since the one-fluid description involves the lowest branch of the spectrum we can neglect the displacement current 
that becomes relevant only for the electromagnetic propagation. For the same reason we can neglect the time 
derivative in Eq. (\ref{onef3}), i.e. $\partial_{t} \vec{J} \ll \Omega_{p}^2 \vec{E}$. Consequently Eqs. (\ref{onef3}) 
and (\ref{MX2}) in the low-frequency branch of the spectrum  become
\begin{equation}
\vec{\nabla}\times \vec{B} = 
4\pi \vec{J} + c_{\omega} \vec{\omega} - c_{B} \vec{B}, \qquad \vec{E} =\vec{J}/\sigma - \vec{v} \times \vec{B}.
\label{MX2a}
\end{equation}
Recalling that $\vec{\nabla} \times \vec{E} = - \partial_{t} \vec{B}$, Eq. (\ref{MX2a}) can be used to obtain an equation that is reminiscent of the magnetic diffusivity equation, namely
\begin{equation}
\partial_{t} \vec{B} = \vec{\nabla}\times(\vec{v} \times \vec{B}) + \frac{\nabla^2 \vec{B}}{4 \pi \sigma} + \frac{1}{4\pi \sigma} \vec{\nabla} \times ( c_{\omega} \vec{\omega}) 
-  \frac{1}{4\pi \sigma} \vec{\nabla} \times ( c_{B} \vec{B}). 
\label{MX2c}
\end{equation}

Introducing the vorticity $\vec{\omega}$ into Eq. (\ref{onef2}) and dividing both sides of the equation by $\rho_{m}$ we obtain 
\begin{equation}
\partial_{t} \vec{v} + \vec{\omega} \times \vec{v} = \frac{\vec{J} \times \vec{B}}{\rho_{m}} - \vec{\nabla} \biggl[ \frac{P}{\rho_{m}} + \frac{v^2}{2} \biggr] + \nu_{kin} \nabla^2 \vec{v}.
\label{v11}
\end{equation}
Taking the curl of Eq. (\ref{v11}) the evolution equation of the vorticity becomes:
\begin{equation}
\partial_{t} \vec{\omega}= 
\vec{\nabla} \times (\vec{v} \times \vec{\omega}) + \frac{\vec{\nabla} \times (\vec{J} \times \vec{B})}{\rho_{m}}  + 
\nu_{kin} \nabla^2 \vec{\omega}.
\label{v12}
\end{equation}
The most interesting solutions of the one-fluid equations will involve the situations where
the vortical and the magnetic currents play the dominant role. However, before turning the attention on these classes  
solutions it is useful to remark that the equilibrium solutions of the plasma at rest (i.e. $\vec{v}=0$)
are simply given by 
\begin{equation}
\vec{\nabla} P = \vec{J} \times \vec{B}, \qquad 4\pi\vec{J} = \biggl(\vec{\nabla}\times \vec{B} - c_{\omega} \vec{\omega} + c_{B} \vec{B} \biggr).
\label{eq}
\end{equation}
From Eq. (\ref{eq}) it is immediate to obtain the following three identities 
\begin{equation}
4\pi \vec{\nabla} P = (\vec{\nabla}\times \vec{B})\times \vec{B}, \qquad (\vec{B}\cdot\vec{\nabla})P =0, \qquad (\vec{J}\cdot\vec{\nabla})P =0.
\label{eq2}
\end{equation}
The relations of Eq. (\ref{eq2}) are not explicitly modified by the presence of vortical and magnetic currents. The two 
last conditions in Eq. (\ref{eq2}) define the so-called magnetic surfaces: the pressure gradient vanishes along the lines of magnetic force 
and along the current lines. We conclude that neither the vortical not the magnetic current affect directly 
the equilibrium solutions.

\subsection{Bulk velocity parallel to the magnetic field direction}
The single fluid AMHD equations admit various solutions that have no counterpart in the case of ordinary MHD. 
Consider first the situation where the hypermagnetic magnetic field and the velocity are parallel and have non-vanishing 
magnetic and kinetic gyrotropy, i.e. 
\begin{equation}
\vec{v} \times \vec{B} =0, \qquad \vec{v} \cdot \vec{\nabla}\times \vec{v} = p_{v}(t) v^2, \qquad  \vec{B} \cdot \vec{\nabla}\times \vec{B} = p_{B}(t) B^2.
\label{J00}
\end{equation}
The simplest way to realize the situation described by Eq. (\ref{J00}) is to require that $\vec{v}$ and $\vec{B}$ 
are both Beltrami-like fields (see appendix \ref{APPB} for this terminology) characterized by 
$\vec{\nabla} \times \vec{v} = p_{v}(t) \vec{v}$ and $\vec{\nabla} \times \vec{B} = p_{B}(t) \vec{B}$.
Moreover, since $\vec{v} \times \vec{B} =0$, it is natural to require that $p_{v}(t) = p_{B}(t)$. From Eq. (\ref{MX2a}) the total current $\vec{J}$ can be easily determined;  the Ohmic electric field is then given by:
\begin{equation}
\vec{E} = \frac{p_{B}(t) + c_{B}(t)}{4 \pi \sigma} \vec{B} - \frac{ c_{\omega}(t)}{4\pi \sigma} \vec{\omega}.
\label{E11}
\end{equation}
From Eq. (\ref{MX2c}) the hypermagnetic field is obtained by solving the following equation
\begin{equation}
\partial_{t} \vec{B} = -\frac{p_{B}(t) [ p_{B}(t) + c_{B}(t)] }{4 \pi \sigma(t)} \vec{B} + \frac{c_{\omega}(t)}{4 \pi \sigma(t)} p_{B}(t) \vec{\omega},
\label{E12}
\end{equation}
where $\vec{\omega}(t,z)$ is the solution of Eq. (\ref{v12}). Thanks to the symmetries of the problem the solution of this equation is given by
$\vec{\omega}(t,z) = \vec{\omega}_{0}(z) \exp{[- \int_{0}^{t} p_{B}^2(t') \nu_{kin}(t') d t']}$, where $\omega_{0}(z)$ is the initial vorticity which 
can also be written as $\vec{\omega}_{0}(z) = p_{B}(0) \vec{v}_{0}(z)$.
Equation (\ref{E12}) can then be solved in general terms. However, recalling that $c_{B}(t)$ and $c_{\omega}(t)$ are explicit functions of time but they 
depend on the rescaled chemical potential\footnote{In this discussion we shall keep the time-dependence in the kinetic coefficients even if, strictly speaking, $c_{B}$ and $c_{\omega}$ may depend on the temperature and of the chemical potential but they are constant in time. However, in curved backgrounds a mild breaking of conformal invariance may induce a time dependence which is, however, not central to the present analysis.} since $p_{B}(t)$ is arbitrary we can choose 
$p_{B}(t) = - c_{B}(t)$. In this case, the solution of Eq. (\ref{E12}) shares the same properties of the general solution but it is 
mathematically simpler:
\begin{equation}
\vec{B}( t, z) = \vec{\omega}_{0}(z) \int_{0}^{t}\,d t^{\prime} \frac{c_{B}(t^{\prime}) c_{\omega}(t^{\prime}) }{4 \pi \sigma(t^{\prime})}  e^{- \int_{0}^{t^{\prime}}c_{B}^2(t^{\prime\prime}) \nu_{kin}(t^{\prime\prime}) d t^{\prime\prime}}.
\label{E13}
\end{equation}
The results of Eq. (\ref{E13}) describe the generation of the hypermagnetic field thanks to some initial vortical current. 
To deepen this question let us assume that $c_{\omega}$ and $c_{B}$ are both constant.
Equation (\ref{E13}) can then be solved and the result is 
\begin{equation}
\vec{B}(z,t) = -\frac{\vec{v}_{0}(z) \, c_{\omega}}{4 \pi \,\nu_{kin} \, \sigma}\biggl[ 1 - e^{ - \nu_{kin} c_{B}^2 t}\biggr],
\label{E14}
\end{equation}
where we used that  $\vec{\omega}_{0}(z) = - c_{B} \vec{v}_{0}(z)$ when $c_{B}$ is constant in time and $c_{B}(0) = c_{B}$.
This result is also valid for a relativistic equation of state (i.e. $w= 4 \rho/3$) provided the incompressible 
closure is consistently adopted and can be easily generalized to curved backgrounds. While these 
generalizations are not germane to our theme it is worth to emphasize that   
in the limit $t \to \infty$ the suppression of the magnetic field is controlled by $4 \pi \sigma \nu_{kin}$ which is 
nothing but the Prandtl number given as the ratio of the magnetic and of the kinetic Reynolds number \cite{biskamp}. 
The Prandtl number is roughly independent on the temperature. For instance at the electroweak epoch \cite{mg19982000} we
 would have that $\nu_{kin} \simeq 1/(\alpha^{\prime\,2}\, T)$ while $\sigma \simeq T/\alpha^{\prime}$
where $\alpha^{\prime}= g^{\prime\, 2}/(4\pi)$. Recalling the results of Eqs. (\ref{eq5}) and (\ref{eq6}) we therefore have that 
\begin{equation}
\lim_{t \to \infty} \vec{B}(t,z) = \frac{c_{\omega}(T)}{4 \pi}\, \alpha^{\prime\,3}\, \vec{v}_{0}, \qquad c_{\omega}(T) = 2 T^2 a(\overline{\mu}_{R}),
\label{E15}
\end{equation}
where $a_{B}(\overline{\mu}_{R})$ is the usual arbitrary function of the rescaled chemical potential $\overline{\mu}_{R} = \mu_{R}/T$ (see Eqs. (\ref{eq5}) and (\ref{eq6})).
The suppression due to the conductivity is therefore eliminated and what is left is a milder suppression ${\mathcal O}(\alpha^{\prime\,3})$. In the same limit the hypermagnetic current turns out to be more suppressed than the vortical current.
As long as the inverse of the Prandtl number scales as 
$\alpha^{\prime\, 3}$ the previous discussion is generally valid and this is what happens in the case of the electroweak plasma \cite{mg19982000} (see also \cite{CC1} for specific 
estimates of the conductivity in the electroweak phase\footnote{Similar kinds of considerations s can also be developed in the case of a strongly interacting 
plasma as long as the same scaling occurs \cite{CC2}.}). 

\subsection{Bulk velocity orthogonal to the magnetic field direction}

If the hypermagnetic field and the bulk velocity of the plasma are orthogonal  (i.e. $\vec{v} \cdot \vec{B} =0$),  employing 
the helical basis of appendix \ref{APPB} the hypermagnetic field and the velocity field  can be written as:
\begin{equation}
\vec{B}(t,z) = B_{0}(t) \hat{z} + B_{1}(t) \hat{a}(z,p), \qquad \vec{v}(z,t) = v(t)  \hat{b}(z,p).
\label{an1}
\end{equation}
From Eq. (\ref{MX2a}) the AMHD current becomes:
\begin{equation}
\vec{J}(t,z) = \frac{B_{1}(t)}{4 \pi} [ c_{B} + p] \hat{a}(z,p) - \frac{c_{\omega}}{4 \pi}\, p\, v(t) \hat{b}(z,p) + \frac{c_{B}B_{0}(t)}{4 \pi} \hat{z}.
\label{an2}
\end{equation}
By analyzing the structure of the evolution equation of the magnetic field and of the vorticity it emerges that 
the system is consistent provided $\partial_{t} B_{0} =0$ (i.e. constant magnetic field along $\hat{z}$) and provided $c_{\omega}=0$. 
In this case the coupled evolution of the vorticity and of the magnetic field obeys 
\begin{equation}
\frac{d \omega}{d t} = \frac{p c_{B} B_{0} }{4 \pi \rho_{m}} B_{1} - \frac{B_{0}}{4 \pi \rho_{m}} p\,[ c_{B} + p] B_{1},\qquad \frac{d B_{1}}{d t} = \omega B_{0} - \frac{p B_{1}}{4 \pi \sigma} [c_{B} + p],
\label{an4}
\end{equation}
where $\omega(t)= p \,v(t)$ and $\omega(z,t)= \omega(t) \hat{b}(z,p)$.
The equations can be diagonalized with a specific choice of the coordinate 
system. The simplest and most convenient one is $p= - c_{B}$; in this case the two equations can be 
combined by differentiating once Eq. (\ref{an4}). The result is 
\begin{equation}
\frac{d^2 B_{1}}{d t^2}  + c_{B}^2 v_{A}^2 B_{1} =0, \qquad v_{A} = \frac{B_{0}}{\sqrt{4 \pi \rho_{m}}},
\label{an5}
\end{equation}
where $v_{A}$ is the Alfv\'en velocity.  This solution has been swiftly presented in Ref. \cite{mg2013} and recently rediscovered in \cite{yamamoto}. Equation (\ref{an5}) describes the AMHD analog of the non-linear Alfv\'en wave. The anomalous Alfv\'en wave 
has been already discussed in section \ref{sec3} as a low-frequency limit of the two-fluid equations. 

\subsection{Fully non-linear evolution and baryogenesis}

So far we considered small fluctuations of the chiral concentration around an otherwise homogeneous 
value denoted by $n_{1}$ in section \ref{sec2}. In the opposite case the AMHD equations imply a specific relation 
between the concentration (or the chemical potential) and the topological properties of the hypermagnetic fields.
To illustrate this point we will show that close to equilibrium the chemical potential is determined not only 
by the magnetic gyrotropy but also by the total vorticity of the plasma. Hypermagnetic field configurations with non-vanishing gyrotropy have been used to model the generation of the baryon or lepton asymmetry \cite{mg19982000} (see also \cite{bb1,bb1a,bb2}).
Consider, therefore, the evolution equation of the chemical potential which can be written as
\begin{equation}
\partial_{t} \overline{\mu}_{R} + \Gamma \overline{\mu}_{R} = - \frac{4 \mu_{0}}{\varsigma} {\mathcal A}_{ R} \vec{E} \cdot \vec{B},
\label{muR}
\end{equation}
where $\Gamma$  is the perturbative rate of he chirality 
flip processes (in the case of \cite{mg19982000} it is determined by the 
scattering of right electrons with the Higgs and gauge bosons and with 
the top quarks because of their large Yukawa coupling). In Eq. (\ref{muR}) we also used the following general relation relation $\overline{\mu}_{R} = \mu_{0} n_{R}/\varsigma$ where $\mu_{0}$ is a numerical factor depending on the specific features of the plasma 
while $\varsigma$ is, as usual, the entropy density. 

To compute $\overline{\mu}_{R}$ in the proximity 
of an equilibrium situation we need to deduce the hyperelectric field. Recalling then Eq. (\ref{MX2a}), the hyperelectric field
ca be related to the total Ohmic current so that Eq. (\ref{muR}) will become
\begin{equation}
\partial_{t} \overline{\mu}_{R} + \Gamma \overline{\mu}_{R} - \frac{8 \,T^2\, a_{B}(\overline{\mu}_{R})\,\mu_{0}}{\sigma \varsigma} \,{\mathcal A}_{R}\, \vec{\omega}\cdot \vec{B}
+ 16 \frac{\mu_{0} \,T \, \overline{\mu}_{R} }{\sigma \varsigma}\,{\mathcal A}_{R}^2\, B^2= - \frac{\mu_{0} }{\pi \sigma \varsigma} \vec{B} \cdot \vec{\nabla} \times \vec{B}.
\label{muR2}
\end{equation}
We shall now choose $a_{B}(\overline{\mu}_{R}) = \overline{\mu}_{R}$; if $a_{B}(\overline{\mu}_{R}) \neq \overline{\mu}_{R}$ in the evolution 
equation of the concentration we should add a further term proportional to $\vec{\nabla}\cdot [ c_{RB}(\overline{\mu}_{R}) \vec{B}]$. This term 
vanishes in the case $a_{B}(\overline{\mu}_{R}) \propto \overline{\mu}_{R}$: $c_{RB}(\overline{\mu}_{R})$ contains the derivative 
of $a_{B}(\overline{\mu}_{R})$ with respect to $\overline{\mu}_{R}$ and it is therefore constant. In more general situations Eq. (\ref{muR2}) will just contain 
a supplementary contribution of the type $\vec{B}\cdot\vec{\nabla}\overline{\mu}_{R}$. 

The rescaled chemical potential enters the infinitely conducting limit (see appendix \ref{APPB}) and the smallness of the particle asymmetries 
is the rationale for the minuteness of the rescaled chemical potentials in approximate thermal equilibrium. At equilibrium, we can determine $\overline{\mu}_{R}$ from Eq. (\ref{muR2}) and the result is given by:
\begin{equation}
\overline{\mu}_{R} = - \biggl(\frac{\mu_{0} {\mathcal A}_{R}}{\pi\, \varsigma\, \sigma}\biggr)\,\,\,\frac{\vec{B} \cdot \vec{\nabla}\times \vec{B}}{[\Gamma + \Gamma_{B} - \Gamma_{\omega}]},
\label{mumu}
\end{equation}
While $\Gamma$ is the perturbative chirality flip rate, the other terms can be understood as rates stemming from the hypermagnetic current and from the vortical current and they are
\begin{equation}
\Gamma_{B} = \frac{16 \mu_{0}}{\varsigma\,\sigma} {\mathcal A}_{R}^2 T B^2, \qquad \Gamma_{\omega} = \frac{8 T^2 \mu_{0}}{\pi \, \sigma\,\varsigma} {\mathcal A}_{R}  \vec{\omega} \cdot \vec{B}.
\end{equation}
In the case of right electrons (see \cite{mg19982000}) ${\mathcal A}_{R} = -
g'^2 y_{R}^2 /(64 \pi^2)$ where $g'$ denotes the gauge coupling and $y_{R} = -2$ is the hypercharge assigment of the right electrons. In the same 
situation we have $\overline{\mu}_{R} = \mu_{0} n_{R}/\varsigma$ and  $\mu_{0} = 87 \pi^2 N_{eff}/220$,
where $N_{eff}$ is the effective number of relativistic degrees of freedom of the system\footnote{There have been a number of suggestions for possible roles that the abelian hypermagnetic Chern-Simons term might play in cosmology. One of them is related to the observation that right-handed electrons, which do not take part in weak interactions and also have a very small Yukawa coupling, are practically decoupled from the thermal ensemble above temperatures of about $10$ TeV.}.
If the plasma is hypercharge neutral  
the value of the chemical potential can be estimated from the asymmetry in the case where all the standard model 
charges are in complete thermal equilibrium. If all the asymmetry is attributed to the right electrons (which is, in some 
sense, the most favourable situation) then $\overline{\mu}_{R} = (87 \pi^2/220) \, N_{\mathrm{eff}} (n_{R}/\varsigma)$ where $N_{\mathrm{eff}} = 106.75$.
 With these specifications Eq. (\ref{mumu}) becomes 
\begin{equation}
\overline{\mu}_{R} = -  \frac{783\, \alpha^{\prime}}{88 \,\pi\, \sigma\, T^3} \frac{\vec{B} \cdot \vec{\nabla}\times \vec{B}}{[\Gamma + \Gamma_{B} - \Gamma_{\omega}]}, \qquad \Gamma_{B} = \frac{783 \, \alpha^{\prime\, 2}}{22 \, \pi^2\, \sigma} \frac{B^2}{T^3},
\label{mumu2}
\end{equation}
where $\alpha' = g^{\prime \,2}/(4\pi)$. Equation (\ref{mumu2}) coincides with the previous results  (see e.g. 
Eq. (6.15) of the last paper of \cite{mg19982000} and see also \cite{bb1,bb1a,bb2}) in the limit $\Gamma_{\omega} \to 0$. The results of Eq. (\ref{mumu2})  show that the final value of the chemical potential depends on the properties of the flow entering the definition of $\Gamma_{\omega}$.
In summary we can say that the magnetic currents and the vortical currents can affect a number of processes such as the formation of the baryon asymmetry or the dynamics of the electroweak phase transition. Similar kinds of effects can be expected in the case of strongly interacting plasmas where the magnetic 
gyrotropy can also determine the properties of the chemical potential.

\newpage
\renewcommand{\theequation}{5.\arabic{equation}}
\setcounter{equation}{0}
\section{Concluding remarks}
\label{sec6}
The dispersion relations of anomalous magnetohydrodynamics are affected by the vortical and the hypermagnetic currents. The vortical currents do not impact on the high-frequency 
branch of the spectrum but the opposite is true at lower frequencies where new solutions describe the simultaneous presence of hypermagnetic knots and fluid vortices. These parity-odd configurations carry, respectively, hypermagnetic and kinetic gyrotropy. 
The physical properties of the system roughly interpolate between the features of conventional chiral liquids and the results valid for cold electromagnetic plasmas. While chiral currents are anomalous and do not contribute to entropy production, vector currents are associated with the generalized Joule heating. 

When chiral and Ohmic currents are simultaneously present the second law of thermodynamics constrains the kinetic coefficients. 
The  hypermagnetic, vortical and Ohmic currents affect the evolution of the gauge fields and determine the hyperelectric field of the plasma.
In anomalous magnetohydrodynamics the perfectly conducting limit is well posed and the hypermagnetic helicity of the knots is strictly conserved, as it happens in the case of conventional plasmas.  The hypermagnetic currents are then completely washed out in the perfectly conducting limit and strongly suppressed when the conductivity is large but finite. Close to thermal equilibrium the concentration of the chiral species and the corresponding chemical potential will depend not only on the hypermagnetic gyrotropy but also and on the vortical currents. 

In summary the evolution equations of anomalous magnetohydrodynamics offer a minimal theoretical framework where the interplay between conduction currents and chiral currents can be quantitatively analyzed. It is therefore fair to say that the results derived here complement and extend some of the present and earlier strategies aimed at an improved understanding of chiral liquids when generalized Ohmic effects cannot be neglected in the evolution of hypermagnetic and hyperelectric fields at finite fermionic density. 
 
\newpage
\begin{appendix}
\renewcommand{\theequation}{A.\arabic{equation}}
\setcounter{equation}{0}
\section{Generalized Appleton-Hartree equation}
\label{APPA}
\subsection{Explicit form of $\varepsilon_{s}$ and $\varepsilon_{v}$}
We are going to give, in what follows the explicit form 
of the dielectric tensors appearing in section \ref{sec3}. The matrix form of $\varepsilon_{s}(\Omega) $ and $\varepsilon_{v}(\Omega) $ 
is given by:
\begin{equation}
\varepsilon_{s}(\Omega) 
= \left(\matrix{\varepsilon_{1}(\Omega)
& i\,\varepsilon_{2}(\Omega) & 0&\cr
-i \varepsilon_{2}(\Omega) & \varepsilon_{1}(\Omega) &0&\cr
0&0&\varepsilon_{\parallel}(\Omega) }\right),\qquad 
\varepsilon_{v}(\Omega) 
= \left(\matrix{\varepsilon_{3}(\Omega)
& -i\,\varepsilon_{4}(\Omega) & 0&\cr
i \varepsilon_{4}(\Omega) & \varepsilon_{3}(\Omega) &0&\cr
0&0&0}\right),
\label{epstenSV}
\end{equation}
where $\varepsilon_{1}(\Omega)$, $\varepsilon_{2}(\Omega)$,  $\varepsilon_{3}(\Omega)$, $\varepsilon_{4}(\Omega)$ and 
$\varepsilon_{\parallel}(\Omega)$ are defined as:
\begin{eqnarray}
&& \varepsilon_{1}(\Omega) = 1 - \frac{\Omega^2_{p\, +} }{\Omega^2- \Omega_{ B\, +}^2} -
\frac{\Omega^2_{p\, -} }{\Omega^2  - \Omega_{B\, -}^2},
\label{eps1}\\
&& \varepsilon_{2}(\Omega) = \biggl(\frac{\Omega_{B\, -}}{\Omega}\biggr) \frac{\Omega^2_{p\,- } }{\Omega^2 - \Omega^2_{B\,-}} 
-  \biggl(\frac{\Omega_{B\, +}}{\Omega} \biggr)\frac{\Omega^2_{p\, +} }{\Omega^2 - \Omega^2_{ B\, +}} .
\label{eps2}\\
&& \varepsilon_{\parallel}(\Omega) = 1 - \frac{ \Omega_{p\,\,+}^2}{\Omega^2} - \frac{ \Omega_{p\,\,-}^2}{\Omega^2},
\label{epspar}\\
&& \varepsilon_{3}(\Omega) = \frac{q \Omega}{(m_{+} +m_{-})} \biggl[\frac{1}{\Omega_{B\,-}^2  - \Omega^2} - \frac{1}{\Omega_{B\,+}^2  - \Omega^2}\biggr],
\label{eps3}\\
&& \varepsilon_{4}(\Omega) = \frac{q}{(m_{+} + m_{-})} \biggl[\frac{\Omega_{B\,-}}{\Omega_{B\,-}^2  - \Omega^2} + \frac{\Omega_{B\,+}}{\Omega_{B\,+}^2  - \Omega^2}\biggr].
\label{eps4}
\end{eqnarray}
Both $\varepsilon_{3}(\Omega)$ and $\varepsilon_{4}(\Omega)$ have dimensions of an inverse frequency squared;
$\varepsilon_{1}(\Omega)$, $\varepsilon_{2}(\Omega)$ and $\varepsilon_{\parallel}(\Omega)$ are instead dimensionless.
The frequencies appearing in Eqs. (\ref{eps1})--(\ref{epspar}) and (\ref{eps3})--(\ref{eps4}) are 
the plasma and the Larmor frequencies associated with the charge carriers of both signs, i.e. 
$\Omega_{p\,\pm} = \sqrt{4 \pi q^2 n_{0}}/m_{\pm}$ and $\Omega_{B\, \pm} = q B_{0}/m_{\pm}$.
To compare the dispersion relations with the standard situation of cold plasmas we must bear in mind that 
the ratios of the plasma and Larmor frequencies are related to the inverse 
ratio of the masses, i.e. $\Omega_{p\,+}/\Omega_{p\, -} =   \Omega_{B\,+}/\Omega_{B\, -} = m_{-}/m_{+}$.

\subsection{The seven function}
The generalized form of the Appleton-Hartree equation (see Eq. (\ref{AH})) depends on $7$ functions whose explicit form is given by:
\begin{eqnarray}
f_{B}(\varepsilon,\, \Omega,\, n,\, \theta) &=& \frac{\varepsilon_{\parallel}}{n^4 \Omega^2} \cos^2{\theta} +  \frac{(\varepsilon_{L} + \varepsilon_{R})}{2 n^4 \Omega^2} \sin^2{\theta},
\nonumber\\
g_{B}(\varepsilon,\, \Omega,\, n,\, \theta) &=& \frac{\varepsilon_{\parallel} (\varepsilon_{L} - \varepsilon_{R})}{n^5 \, \Omega}\cos{\theta},
\nonumber\\
f_{\omega}(\varepsilon,\, \Omega,\, n,\, \theta) &=& \frac{\cos^2{\theta}}{n^4} \biggl\{ (\varepsilon_{3} - \varepsilon_{4}) (\varepsilon_{3} + \varepsilon_{4}) \varepsilon_{\parallel} + n^2 \biggl[\sin{\theta} \varepsilon_{3}^2- ( \varepsilon_{3}^2 - 2\varepsilon_{4}^2)\sin^2{\theta}\biggr]\biggl\},
\nonumber\\
g_{\omega}(\varepsilon,\, \Omega,\, n,\, \theta) &=& \frac{\cos{\theta}}{2 n^5} \biggl\{2 \varepsilon_{\parallel}\biggl[ - n^2 \varepsilon_{4}+ \varepsilon_{3} (\varepsilon_{L} - \varepsilon_{R}) + \varepsilon_{4} (\varepsilon_{L} + \varepsilon_{R})\biggr] - 2 n^2 \varepsilon_{4} \varepsilon_{\parallel} \cos^2{\theta}  
\nonumber\\
&+& n^2 \varepsilon_{3} (\varepsilon_{L} - \varepsilon_{R}) \sin{\theta} + n^2\biggl[ 4 n^2 \varepsilon_{4} - 2 \varepsilon_{3} (\varepsilon_{L} - \varepsilon_{R}) - 3 \varepsilon_{4} (\varepsilon_{L} + \varepsilon_{R}) \biggr]\sin^2{\theta}\biggr\},
\nonumber\\
h_{1}(\varepsilon,\, \Omega,\, n,\, \theta) &=& - \frac{2 \varepsilon_{4} }{ n^3 \Omega^2} \cos{\theta} \sin^2{\theta},
\nonumber\\
h_{2}(\varepsilon,\, \Omega,\, n,\, \theta) &=& \frac{1}{2 n^4 \Omega} \biggl\{ 4 \varepsilon_{3} \varepsilon_{\parallel} \cos^2{\theta} + \sin{\theta}\biggl[ \varepsilon_{3} (\varepsilon_{L} + \varepsilon_{R}) + \varepsilon_{4} (- \varepsilon_{L} + \varepsilon_{R}) \sin{\theta}\biggr]
\biggr\},
\nonumber\\
h_{3}(\varepsilon,\, \Omega,\, n,\, \theta) &=&- \frac{\varepsilon_{3} \, \varepsilon_{4} \sin{2 \theta} ( 1 + \sin{\theta})}{2\,n^3 \Omega}.
\label{seven}
\end{eqnarray}
The functions reported in Eq. (\ref{seven}) determine, through Eq. (\ref{AH}), the form of the dispersion relations when the hypermagnetic and the vortical currents are simultaneously present in the anomalous magnetohydrodynamics equations.  

\renewcommand{\theequation}{B.\arabic{equation}}
\setcounter{equation}{0}
\section{Hypermagnetic knots and Beltrami fields}
\label{APPB}
In the resistive approximation, the hyperelectric and the hypermagnetic fields are not exactly orthogonal and the nature 
of this misalignment is crucial both for the generation of the baryon asymmetry and for the chiral magnetic effect. 
In AMHD the induced hyperelectric field stems directly from the approximate form of the Ohm's law and it vanishes exactly, in the 
plasma frame, when the conductivity goes formally to infinity.  In the same limit the 
contribution of the chemical potential to the anomalous hypermagnetic diffusivity equation gets always 
erased. At finite conductivity the anomalous contribution can be often rephrased in terms of the magnetic gyrotropy \cite{biskamp}
which defines hypermagnetic knot solutions \cite{mg19982000}. 
\subsection{Hypermagnetic knots}
The configurations minimizing the hypermagnetic 
energy density with the constraint that the helicity be conserved coincide, in the perfectly conducting limit, with the ones obtainable in 
ideal magnetohydrodynamics where the anomalous currents are neglected \cite{moffat,biskamp,alf,parker}. 

In the perfectly conducting limit Eq. (\ref{MX2c}) leads to 
$\partial_{t} \vec{B} = \vec{\nabla}\times(\vec{v} \times \vec{B}) + {\mathcal O}(\overline{\mu}_{R}/\sigma)$
which is qualitatively similar to the result of conventional magnetohydrdynamics. 
Defining the vector potential in the Coulomb gauge, the magnetic diffusivity equation becomes, up to small corrections, 
$\partial_{t} \vec{A}= \vec{v} \times (\vec{\nabla}\times\vec{A})$. The analysis 
of Ref. \cite{woltjer} can then be exploited. The magnetic energy density shall then be minimized in a finite volume 
under the assumption of constant magnetic helicity by introducing the Lagrange multiplier $p_{B}$. By taking the functional variation 
of ${\mathcal G} =\int_{V} d^{3} x\{ |\vec{\nabla} \times \vec{A}|^2 - p_{B} \vec{A} \cdot (\vec{\nabla}\times\vec{A}) \}$,
with respect to $\vec{A}$ and by requiring $\delta {\mathcal G} =0$, the 
configurations extremizing ${\mathcal G}$ are such that $\vec{\nabla} \times \vec{B} = p_{B} \vec{B}$.  In performing the functional variation we assumed that $V$ is the fiducial volume of a closed system.  

The configurations $\vec{\nabla} \times \vec{B} = p_{B} \vec{B}$ have been used to describe hypermagnetic knots (see \cite{mg19982000}, third and fourth papers); in this case $q$ has dimensions of an inverse length and sets the scale of the hypermagnetic knot which is related to Chern-Simons waves.  The configurations with constant $p_{B}$ represent the lowest state of magnetic energy which a closed system may attain also in the case where anomalous currents are present, provided the ambient plasma is perfectly conducting.

\subsection{Gyrotropic bases}
The knotted solutions can be expanded in an appropriate gyrotropic basis. Let us then consider a vector field $\vec{a}$ fields satisfying 
$\vec{a} \times (\vec{\nabla} \times \vec{a})=0$. The simplest realization of these  Beltrami fields is provided 
by the eigenvectors of the curl operator but more general situations are know and have been 
extensively examined in the literature. Two gryrotropic and orthonormal bases of opposite parity are given by 
 ($\hat{a}$, $\hat{b}$, $\hat{z}$) and by ($\hat{c}$, $\hat{d}$, $\hat{z}$)
\begin{eqnarray}
&& \hat{a}(z,p) = \{ \cos{p z},\, - \sin{p z}, 0\}, \qquad \hat{b}(z,p) = \{ \sin{p z},\,  \cos{p z}, 0\},
\label{basis1}\\
&& \hat{c}(z,p) = \{ \cos{p z},\,  \sin{p z}, 0\}, \qquad \hat{d}(z,p) = \{ -\sin{p z},\,  \cos{p z}, 0\},
\label{basis2}
\end{eqnarray}
As anticipated the bases of Eqs. (\ref{basis1}) and (\ref{basis2}) are orthonormal. Indeed we have
$\hat{a}\cdot\hat{b} = \hat{a}\cdot\hat{z} = \hat{b}\cdot\hat{z} =0$ and  $(\hat{a} \times \hat{b}) \cdot\hat{z} =1$
(and similarly for $\hat{c}$, $\hat{d}$ and $\hat{z}$). The unit vectors 
of Eqs. (\ref{basis1}) and (\ref{basis2}) are normalized eigenvectors of the curl operator 
with eigenvalues $+ p$ and $-p$. 

In ordinary MHD knotted solutions can be constructed from Beltrami fields by postulating a solenoidal (static) 
current and by neglecting the displacement current. In anomalous magnetohydrodynamics these simple constructions cannot be 
immediately extended because of the magnetic and vortical currents.
The knot solutions obtainable by extremizing the functional ${\mathcal G}$
correspond to uniform magnetic fields well inside the core of the knot. This conclusion is 
evident if we use the basis of Eqs. (\ref{basis1}) and (\ref{basis2}). For instance in the limit $p z <1$ the field configuration 
$\vec{B}(z, p) = B_{0} \hat{a}(z, p)\to B_{0} \hat{x} $ is practically uniform and directed along the $\hat{x}$ axis. The connections between 
Beltrami fields, force-free solutions in ordinary MHD equilibrium and electromagnetic waves propagation 
have been explored in a number of papers \cite{woltjer,kendall,chu,saling} starting from the classic works 
of Fermi and Chandrasekhar \cite{fermi}. It is also possible to obtain hypermagnetic knot solutions 
with finite helicity and finite gyrotropy which do not satisfy the relation of Beltrami fields. These 
solutions have been studied in a number of interesting frameworks  (see last two papers of Ref. \cite{mg19982000} and also \cite{HK2,HK3}).

\end{appendix}


\newpage

\begin{thebibliography}{99}

\bibitem{moffat} H. K. Moffat, {\it Magnetic field generation in electrically conducting fluids}, (Cambridge University Press, 
Cambridge 1978).

\bibitem{biskamp} D. Biskamp, {\it Magnetohydrodynamic Turbulence}, (Cambridge University Press, 2003).

\bibitem{alf} H. Alfv\'en and C.-G. F\"althammer, {\it Cosmical Electrodynamics}, 2nd edn., (Clarendon press, Oxford, 1963).

\bibitem{parker} E. Parker, {\it Cosmical Magnetic Fields} (Oxford University Press, Oxford, 1979).

\bibitem{zeldovich} Ya. B. Zeldovich, A. A. Ruzmaikin, and D. Sokoloff {\it Magnetic Fields in Astrophysics} (Gordon and Breach, New York 1983).

\bibitem{stix} T. H. Stix, {\it Waves in Plasmas} (American Institute of Physics, 1992).
  
\bibitem{mg2013} M.~Giovannini, Phys.\ Rev.\ D {\bf 88}, 063536 (2013).  

\bibitem{rub} V.A. Rubakov, Prog. Theor. Phys. {\bf 75}, 366 (1986); 
V.A. Matveev {\it et al.}, Nucl. Phys. B {\bf 282}, 700 (1987); V.A. Rubakov, A.N. Tavkhelidze, Phys. Lett. B {\bf 165}, 109 (1985).

\bibitem{red}  A. N. Redlich and L. C. R. Wijewardhana, Phys. Rev. Lett. {\bf 54}, 970 (1985).

  \bibitem{mg19982000} M.~Giovannini and M.~E.~Shaposhnikov,
  Phys.\ Rev.\ D {\bf 57}, 2186 (1998); M.~Giovannini and M.~E.~Shaposhnikov,
  Phys.\ Rev.\ Lett.\  {\bf 80}, 22 (1998); M.~Giovannini,
  Phys.\ Rev.\ D {\bf 61}, 063502 (2000); Phys.\ Rev.\ D {\bf 61}, 063004 (2000).

 \bibitem{axion1} R. D. Peccei and H. R. Quinn, Phys. Rev. Lett. {\bf 38}, 1440  (1977); Phys. Rev. D {\bf 16}, 1791 (1977).

\bibitem{axion2} J. Kim, Phys. Rep. {\bf 150}, 1 (1987); H.-Y. Cheng, {\em ibid}., {\bf 158}, 1 (1988); G. G. Raffelt, Phys. Rep. {\bf 198}, 1 (1990);  Lect.\ Notes Phys.\  {\bf 741}, 51 (2008).
   
\bibitem{ax3} S. Carroll, G. Field and R. Jackiw, Phys. Rev. D {\bf 41},  1231 (1990);  W. D. Garretson, G. Field and S. Carroll, Phys. Rev. D {\bf 46}, 5346 (1992); G. Field and S. Carroll Phys.Rev.D {\bf 62}, 103008 (2000).

\bibitem{kh} D. Kharzeev, L. McLerran and H. Warringa, Nucl. Phys. A {\bf 80}, 3227 (2008); 
K.~Fukushima, D.~Kharzeev and H.~Warringa, Phys.\ Rev.\ D {\bf 78}, 074033 (2008);
D.~Kharzeev,  Annals Phys.\  {\bf 325}, 205 (2010).

\bibitem{hol} K.~Landsteiner, E.~Megias, L.~Melgar and F.~Pena-Benitez,
  Fortsch.\ Phys.\  {\bf 60}, 1064 (2012);  K.~Landsteiner, E.~Megias, L.~Melgar and F.~Pena-Benitez,  JHEP {\bf 1109}, 121 (2011); V.~A.~Rubakov,
  arXiv:1005.1888 [hep-ph]; T. Kalaydzhyan and I. Kirsch, Phys. Rev. Lett. {\bf 106}, 211601 (2011); I. Gahramanov, T. Kalaydzhyan, and I. Kirsch, Phys. Rev. D {\bf 85}, 126013 (2012); 
  V.P. Nair, R. Ray, and S. Roy, Phys. Rev. D {\bf 86}, 025012 (2012).  

\bibitem{SS}  D.~T.~Son and P.~Surowka,  Phys.\ Rev.\ Lett.\  {\bf 103}, 191601 (2009).

\bibitem{wop} H. L. P\'ecseli, {\it Waves and oscillations in plasmas} (CRC press, Boca Raton 2012).

\bibitem{CC1} J.~Ahonen, Phys.\ Rev.\ D {\bf 59}, 023004 (1999); J.~Ahonen and K.~Enqvist,  Phys.\ Lett.\ B {\bf 382}, 40 (1996); H. Heiselberg, Phys. Rev. D {\bf 49}, 4739 (1994).

\bibitem{CC2} Y. Hirono, M. Hongo, and T. Hirano, Phys. Rev. C  {\bf 90}, 021903 (2014); Y. Yin, Phys. Rev. C {\bf 90}, 044903 (2014).

\bibitem{chu} C. Chu and T. Ohkawa, Phys. Rev. Lett. {\bf 48},  837 (1982).

\bibitem{saling} N. Salingaros, Phys. Rev. A {\bf 35}, 4856 (1986); {\it ibid}. {\bf 45}, 8811 (1992); {\it ibid}. {\bf 45}, 8816 (1992).

\bibitem{woltjer} S. Chandrasekhar and L. Woltjer, Proc. Natl.  Acad. Sci.  USA {\bf 44}, 285 (1958); L. Woltjer, Proc. Natl.  Acad. Sci.  USA {\bf 44}, 489 (1958).

\bibitem{kendall} S.Chandrasekhar and P. C. Kendall, Astrophys. J. {\bf 126}, 457 (1976).

\bibitem{yamamoto}  N.~Yamamoto,  arXiv:1505.05444 [hep-th].

\bibitem{bb1} G. Piccinelli and A. Ayala,  Lect. Notes Phys. {\bf 646} (2004);  K.~Bamba,  Phys.\ Rev.\ D {\bf 74}, 123504 (2006); L.~Campanelli and M.~Giannotti,
  Phys.\ Rev.\ Lett.\  {\bf 96}, 161302 (2006).
  
\bibitem{bb1a} K.~Bamba, C.~Q.~Geng and S.~H.~Ho,  Phys.\ Lett.\ B {\bf 664}, 154 (2008); L.~Campanelli,
Int.\ J.\ Mod.\ Phys.\ D {\bf 18}, 1395 (2009); L.~Campanelli,  Eur.\ Phys.\ J.\ C {\bf 74}, 2690 (2014); S.~Ozonder,
  Phys.\ Rev.\ C {\bf 81}, 062201 (2010) [Phys.\ Rev.\ C {\bf 84}, 019903 (2011)].
  
\bibitem{bb2} B.~A.~Fayzullaev, M.~M.~Musakhanov, D.~G.~Pak and M.~Siddikov,
  Phys.\ Lett.\ B {\bf 609}, 442 (2005); M.~N.~Chernodub and A.~J.~Niemi,
  Phys.\ Rev.\ D {\bf 79}, 077901 (2009);  P.~M.~Akhmet'ev, V.~B.~Semikoz and D.~D.~Sokoloff,
  Pisma Zh.\ Eksp.\ Teor.\ Fiz.\  {\bf 91}, 233 (2010).

\bibitem{fermi} E. Fermi and S. Chandrasekhar, Astrophys.J. {\bf 118}, 116 (1953).

\bibitem{HK2} C.~Adam, B.~Muratori and C.~Nash,
  Phys.\ Rev.\ D {\bf 61}, 105018 (2000); R.~Jackiw and S.~Y.~Pi,
  Phys.\ Rev.\ D {\bf 61}, 105015 (2000).  
 
 \bibitem{HK3}  C.~Adam, B.~Muratori and C.~Nash, Phys.\ Rev.\ D {\bf 62}, 105027 (2000);
  Phys.\ Lett.\ B {\bf 485}, 314 (2000).

\end{thebibliography}
\end{document}